\documentclass[aps,prd,twocolumn,showpacs,superscriptaddress,nofootinbib]{revtex4}  
\usepackage{graphicx}  
\usepackage{dcolumn}   
\usepackage{bm}        
\usepackage{amssymb}   
\usepackage{setspace}
\usepackage{hyperref}
\usepackage{textcomp}
\usepackage{amsmath}
\usepackage{palatino}
\usepackage{amsfonts}
\begin{document}

\title{Behavior of the Universe anisotropy in a Big-Bounce cosmology}

\author{Riccardo Moriconi}
\email{moriconi@na.infn.it}
\affiliation{Dipartimento di  Fisica "E. Pancini", Universit\`{a} di Napoli "Federico II",
 Compl. Univ. di Monte S. Angelo, Edificio G, Via Cinthia, I-80126, Napoli, Italy,}
\affiliation{Istituto Nazionale di  Fisica Nucleare (INFN) Sez. di Napoli, Compl. Univ. di Monte S. Angelo,
Edificio G, Via Cinthia, I-80126, Napoli, Italy,}

\author{Giovanni Montani}
\email{giovanni.montani@frascati.enea.it}
\affiliation{Dipartimento di Fisica (VEF), P.le A. Moro 5 (00185) Roma, Italy,}
\affiliation{ENEA, Unità Tecnica Fusione, ENEA C. R. Frascati, via E. Fermi 45, 00044 Frascati (Roma), Italy,}

\date{\today}

\begin{abstract}
We investigate the classical and quantum behavior of a Bianchi I model in the
presence of a stiff matter contribution when the Vilenkin interpretation of the wave function of the Universe is taken into account. We study its
evolution in the so-called polymer
representation of quantum mechanics, in order to characterize the modifications that a discrete nature in the
isotropic variable of the Universe induces on the morphology of the cosmological singularity. 
We demonstrate that in such a model the Big-Bang 
singularity is removed at a semiclassical level in favor of a Big-Bounce when a lattice on the isotropic variable is considered. 
Furthermore, the analysis of the mean values on the quantum degrees of freedom, \textit{i.e} the variables $\beta_{+},\beta_{-}$ in the Misner picture, and the investigation on the evolution of the wave packets show how the typical diverging behavior associated to the anisotropies of the Universe in proximity of the initial singularity disappears in our polymer modified scheme. Indeed, the anisotropies remain finite across the Big-Bounce and they assume a value that depends on the initial conditions fixed far from the turning point.
Finally, we demonstrate that the proposed scenario can be extended, with a suitable choice of the configuration parameters, to the Bianchi IX cosmology and therefore it can 
be regarded as a paradigm for the generic cosmological model. 

\end{abstract}
\maketitle

\section*{INTRODUCTION}

The Bianchi cosmological models are the simplest generalization of  the isotropic Universe and they are possible candidate to describe the nature  of the initial singularity \cite{BKL},\cite{landau},\cite{primordial}. 

Two Bianchi model are of particular dynamical interest, the type I, because it contains the metric time derivatives dominating near the singularity, and the type IX, which generalizes the closed Robertson-Walker dynamics and thus having a space curvature, responsible for a chaotic behavior near the singularity \cite{gravitation}.

Indeed, the Bianchi types VIII and IX (however the type VIII does not admit  an isotropic limit) are the most general models allowed by the homogeneity and their chaotic features are typical of the generic inhomogeneous cosmological solution. 

The anisotropic components of the metric tensor can be easily separated for the Bianchi models from the isotropic component, which is associated to the Universe volume and this separation takes place in a very elegant representation, by using the so-called Misner variables \cite{misnermixmaster}.

Near the initial singularity, the Universe volume vanishes and the Bianchi model anisotropies typically diverges. 

However, it is a common belief that the cosmological singularity must be replaced by a Big-Bounce of the Universe, \textit{i.e.} the volume reaches a minimum and then re-expands. This picture has been reproduced in many quantum approaches to the early Universe dynamics, especially in the so-called Loop Quantum Cosmology \cite{loop1},\cite{loop2},\cite{loop3}. Such a reliable prediction of the quantum cosmology has given rise to a new theoretical framework to interpret the Universe history, dubbed Big-Bounce Cosmology \cite{brandenberger},\cite{bigbounce},\cite{pittorino}. In fact, some important aspects of the Universe dynamics, especially in its early stages, like the basic paradoxes of the Standard cosmological Model \cite{kolb}, could be differently addressed in view of the pre-Big-Bounce history. 

A crucial point in the direction of a revised point of view on the primordial Universe, once the existence of a Big-Bounce is postulated, requires a precise understanding of the role played by the anisotropies degrees of freedom near the primordial turning point of the Universe: by other words, which is the behavior of the Universe anisotropies across the  Big-Bounce. The most natural arena in which testing such a feature of the primordial cosmology is given by the Bianchi model, especially the type I and IX respectively. In this paper, we address exactly this problem, by reproducing the Big-Bounce via a Polymer quantum approach \cite{corichi},\cite{corichidue}. However, to provide a clear physical meaning to the anisotropy variable wavefunction, we combine the polymer technique with the original Vilenkin semiclassical approach to the Wheeler-DeWitt equation \cite{vilenkin}. More specifically, we describe the Bianchi models via the Misner variables \cite{misner}and we retain the volume as a quasi-classical coordinate, although obeying to the modified Hamilton-Jacobi equation, due to the polymer discretization. The anisotropies Misner variables are instead treated as pure quantum degrees of freedom. 

The main technical merit of the present analysis is just to reconcile two different, but complementary points of view, as the mentioned above, in order to provide a consistent and regularized Big-Bounce cosmology, in which the behavior of the anisotropies near and far from the turning point can be discussed in detail. 

We observe how, the dynamics of the semi-classical Universe volume is analyzed in the presence of stiff matter. This choice actually underlying technical reasons, but it is also cosmologically relevant since the stiff matter equation of state mimic the dynamics of a free massless scalar field, whose energy fills the Universe and determines its evolution: this is just the case of an inflaton field at sufficiently high temperature,  where the potential term, responsible for the inflation scenario, can be neglected (actually the cut-off energy density is supposed to be at the Planck scale, well above with respect to the inflation threshold).   

We study in detail the behavior of the wave packets, comparing it with the prediction of the Ehrenfest theorem \cite{ehrenfest}. We recall that the Vilenkin approach allows to deal with an ordinary Schroedinger equation for the anisotropy variables, although the translation of the dynamical picture in the synchronous reference involves the details of the polymer Hamilton-Jacobi equation, describing the leading order dynamics of the Universe volume.

We first address a careful analysis of the Bianchi I model, demonstrating that the anisotropies mean values and variances remain finite near the singularity, differently from the Einsteinian classical behavior, associated to a divergence of the Universe anisotropy near the singularity. For other analyses reaching the same conclusions in the context of the Pre Big-Bang scenario in String theory and in the presence of a scalar field with equation of state $w \geq 1$ (where $w$ is the effective squared sound velocity), see \cite{veneziano},\cite{erickson} respectively.

As far as the WKB Vilenkin approximation holds, we can claim that the anisotropy of the Universe remains finite across the Big-Bounce, its limit valued being dependent on the initial conditions, fixed far from the turning point. 
It is worth noting that WKB assumption is reliable since the Universe volume is essentially a time-like variable in the minisuperspace, more than a real physical degree of freedom.
The anisotropy variables behavior tends also to become more and more classical near the Big-Bounce, in the sense that the ratio between the variance to the mean value decreases. This feature could be also recovered in a non-polymer representation of the Universe volume quasi-classical dynamics, but, in that case, it remains meaningless due to the intrinsic divergence of the anisotropies mean value, as we discuss in some detail.

We stress that in Section \ref{sec:3} and Section \ref{sec:4} we analyze the same dynamical features of the Bianchi I model without and with the polymer paradigm. The comparison of the results outlined in such sections allows to understand the most significant modifications due to the cut-off physics: i) the appearance of a big bounce and ii) the finite asymptotic behavior of the anysotropies.

Then, we extend the same analysis to the Bianchi IX model, estimating the relative behavior of the kinetic term of the Hamiltonian (present in both the Bianchi I and IX types) versus the potential term of the Bianchi IX model, due to its non-vanishing spatial curvature(for Bianchi I the three-dimensional Ricci scalar identically vanishes). 

This study demonstrate that, for a non-zero set of initial conditions (fixed far from the Big-Bounce), the potential term is negligible in the Bianchi IX dynamics, with respect to the kinetic one: all the considerations developed for the Bianchi I model can be applied to the Bianchi IX one too. We estimate the behavior of the potential term on the mean values of the anisotropic variables, but, their increasing classical behavior to the Big-Bounce ensures the predictivity of this quantity toward the average value of the potential (rigorously involved in the Ehrenfest theorem). 

The study of the Bianchi IX dynamics and the possibility to claim a regular behavior of the Universe anisotropies across the Bounce, provides our results with a an high degree of generality, since the Bianchi IX model is the homogeneous prototype for the generic inhomogeneous cosmological solution \cite{BKL1982},\cite{primordial}. The most relevant conceptual progress contained in the present analysis concerns the possibility, on a different regularized framework, to elucidate the original result of Misner about the existence of semi-classical states off the Universe anisotropies, near enough to the singularity \cite{misner}. The Misner conclusion involves only high occupation numbers of the anisotropy degrees of freedom and its physical interpretation remains obscure, due to the diverging character of the anisotropy variables according to the Ehrenfest theorem. The proper interpretation of that Misner result comes out, as soon as, a Big-Bounce and WKB cosmology is implemented for the Universe volume. 

The structure of the paper is organized as follows.

In Section \ref{sec:1} we introduce the Vilenkin approach for the description of the wave function of the Universe, in which is illustrated, for a generic homogenous universe, how to define a genuine definition of semi-positive probability.

Then, in Section \ref{sec:2}, the polymer representation of the quantum mechanics is considered for a simple one-dimensional particle system. In particular, we introduce the basic concepts of the theory as the kinematical and the dynamical proprieties and we will highlight the modified effective Hamiltonian that a polymer modification induces.

The Section \ref{sec:3} is dedicated to the application of the Vilenkin approach to a specific cosmological model: the Bianchi I model in presence of a stiff matter. Firstly, we faced the semiclassical dynamics by studying the Hamilton-Jacobi equation and evaluating the evolution of the variable related to the Universe volume near the initial singularity. Then, the WKB expansion due to the Vilenkin form of the wave function, allow to describe the behavior of the pure quantum degrees of freedom of the system: the anisotropies. For the description of such a variables, a Schrodinger-like equation is resemble whose solution represents the quantum states of the system. Starting from this states it is possible to build the wave packets associated to the wave function of the Universe and compare their evolution with the trajectories of the mean values obtained from the Ehrenfest theorem. Furthermore, both from the semiclassical and the quantum point of view, is stressed the equivalence in the obtained results making the two different polarization choices for the isotropic component of the wave function.

Moreover, the polymer generalization of the latter model is illustrated in Section \ref{sec:4}. In particular, we consider the modifications induced in the configuration variables dynamics towards the initial singularity when a polymer discretization of the Universe volume occurs.

The polymer modification applied to the Bianchi I model is then extended to the more general Bianchi IX model in Section \ref{sec:5}. The main focus in this section is devoted to the importance of the curvature potential term of the Bianchi IX model next to the Big-Bounce.

Brief concluding remarks complete the paper.

\section{Interpretation of the wave function of the Universe}
\label{sec:1}
The behavior of the Universe in quantum cosmology is described by the wave function of the Universe $\psi$\cite{HH}, which represents the solution of the Wheeler-de Witt (WDW) equation\cite{dewitt1},\cite{dewitt2},\cite{dewitt3}. One of the main issues related to the wave function of the Universe is its probabilistic interpretation.
In a generic quantum mechanics system described by a wave function $\psi(x_{i},t)$, where $x_{i}$ are coordinates and $t$ is the time, the probability to find the system in a particular configuration space element $d\Omega_{x}$ is:
\begin{equation}
\label{prob1}
dP = |\psi(x_{i},t)|^{2}d\Omega_{x}.
\end{equation}
The definition above provides in any case a positive semidefinite probability, $dP \geq 0$, and a well-normalized system:
\begin{equation}
\label{norm1}
\int |\psi(x_{i},t)|^{2}d\Omega_{x} = 1.
\end{equation}
In quantum cosmology the wave function of the Universe, defined on the superspace\cite{superspace}, depends on the configuration of the three-metrics $h_{ij}(x)$ and the matter fields $\phi(x)$ without an explicit time dependence.
To discuss the problem in a simple way, let us consider the homogenous minisuperspace models\cite{ADM},\cite{landau},\cite{gravitation}, in which the three-metrics and the matter fields does not depend on the position $x$. The action for this class of model is
\begin{equation}
\label{azione1}
S = \int dt \{p_{\alpha}\dot{h}^{\alpha} - N[h^{\alpha \beta}p_{\alpha}p_{\beta} + U(h)]\},
\end{equation}
where $h^{\alpha}$ represent the superspace variables, $p_{\alpha}$ are the conjugated momenta to $h^{\alpha}$, $N = N(t)$ is the lapse function, $h^{\alpha \beta}$ is the superspace metric and $U(h)$ takes into account the spatial curvature and the potential energy of matter field.

If we decide to proceed in analogy with the Eq.(\ref{prob1}), a straightforward extension for the probability is:
\begin{equation}
\label{prob2}
dP = |\psi(h_{\alpha})|^{2}\sqrt{h}d^{n}h.
\end{equation}
The problem with the definition (\ref{prob2}) is the ``time'' dependence of the variables $h_{\alpha}$. The consequence is that the probability  is not normalizable. In fact, taking into account the term $\sqrt{h}d^{n}h$ is equivalent to consider, in a generic quantum mechanics systems, the quantity $d\Omega_{x}dt$. For the latter case we have
\begin{equation}
\label{norm2}
\int |\psi(x_{i},t)|^{2}d\Omega_{x}dt = \infty,
\end{equation}
and proceeding by analogy in quantum cosmology we have
\begin{equation}
\label{norm3}
\int |\psi(h_{\alpha})|^{2}\sqrt{h}d^{n}h = \infty.
\end{equation}
A way to avoid this consists to provide an alternative definition of probability based on the conserved current
\begin{equation}
\label{corrente1}
J^{\alpha} = -\frac{i}{2}h^{\alpha\beta}[\psi^{*}\nabla_{\beta} \psi - \psi\nabla_{\beta} \psi^{*}] \quad , \quad \nabla_{\alpha} J^{\alpha} = 0.
\end{equation}
This way, the probability to find the Universe in a particular state is 
\begin{equation}
\label{prob3}
dP = J^{\alpha}d \Sigma_{\alpha},
\end{equation}
where d$\Sigma_{\alpha}$ represents the separation between the three-dimensional surfaces on which the current is defined. These surfaces play a role similar to that of constant-time surfaces in conventional quantum mechanics. Furthermore, if we consider the conservation of the current (\ref{corrente1}), than the conservation of the probability is ensured.

The problem with the probability (\ref{prob3}) is that it can be negative, as it is easy to show if you consider a given wave function $\psi$ and then the complex-conjugate $\psi^{*}$. The situation is the same that happens for the negative probabilities in the Klein-Gordon equation (the WDW equation resembles exactly a Klein-Gordon equation with a variable mass).  

The Vilenkin interpretation of the wave function of the Universe\cite{vilenkin} appears as a solution to solve this issue. Such approach consists in the separation, for the configuration variables, in two classes: \textit{semiclassical} and \textit{quantum}. Following this prescription, the quantum variables represent a small subsystem of the Universe and the semiclassical variables act as an external observer for the quantum dynamics, or in other words the effects of the quantum variables on the semiclassical ones are negligible.
We choose to describe for the configuration space the notation $q_{\alpha}$ for the semiclassical variables and $\rho_{\nu}$ for the quantum variables.
The WDW equation for the action (\ref{azione1}) takes the form
\begin{equation}
\label{WDW1}
(\nabla^{2} - U - H_{\rho}) \psi = 0 \quad , \quad \nabla^{2} = \frac{1}{\sqrt{h}}\partial_{\alpha}(\sqrt{h}h^{\alpha \beta}\partial_{\beta})\psi,
\end{equation}
where $h = |\det h_{\alpha \beta}|$ and $\partial_{\alpha} = \frac{\partial}{\partial h^{\alpha}}$.
The operator $\nabla^{2} - U$ that appears in the WDW equation, is the part that survives when we neglect all the quantum variables $\rho_{\nu}$ and their conjugated momenta. For this reason, the other part $H_{ \rho}$ is due to the presence of the quantum subsystem and its smallness is ensured by the existence of a small parameter $\epsilon$ for which 
\begin{equation}
\frac{H_{ \rho} \psi}{(\nabla^{2} - U)\psi} = O(\epsilon),
\end{equation}
where $\epsilon$ is a small parameter proportional to $\hbar$. Also the superspace metric can be expanded in terms of $\epsilon$ as
\begin{equation}
h_{\alpha \beta} = h_{\alpha \beta}^{0}(q) + O(\epsilon)
\end{equation}
and the wave function of the Universe can be written as
\begin{equation}
\label{funz onda}
\psi = A(q) e^{i S(q)} \chi(q,\rho).
\end{equation}
In order to perform a WKB expansion as an expansion series in $\epsilon$ in a properly way, we point out that the potential term $U(q)$ is of the order $\epsilon^{-2}$ and the action $S(q)$ is of the order $\epsilon^{-1}$. This way, if we consider the wave function (\ref{funz onda}) in the Eq.(\ref{WDW1}) we obtain, at the lowest order in $\epsilon$, the Hamilton-Jacobi equation for $S$:
\begin{equation}
\label{H-J 1}
h_{\alpha \beta}\nabla_{\alpha}S\nabla_{\beta}S + U = 0.
\end{equation}
At the next order we obtain the Equation:
\begin{equation}
\label{quant 1}
2 \nabla A\nabla S + A \nabla^{2}S + 2i \nabla S \nabla \chi - H_{\rho}\chi = 0.
\end{equation}

The terms of the Eq.(\ref{quant 1}) can be decoupled in a pair of equations making use of the \textit{Adiabatic Approximation}\cite{adiabatic}. It consists in requiring that
the semiclassical evolution be principally contained in the semiclassical part of the wave function, while the quantum part depends on it only parametrically. The adiabatic approximation is therefore expressed by the condition
\begin{equation}
\label{app adia}
|\partial_{q} A(q)| \gg |\partial_{q} \varphi(q,\rho)|.
\end{equation}
Using the relation (\ref{app adia}) in the Eq.(\ref{quant 1}) we obtain that:
\begin{equation}
\label{deco}
\frac{1}{A} \nabla(A^{2}\nabla S) = 0 \quad , \quad 2i \nabla S \nabla \chi - H_{\rho}\chi = 0.
\end{equation}

The first equation represents  the conservation of the current defined in Eq.(\ref{corrente1}) obtained neglecting the quantum part of the wave function, or in other words using a wave function 
\begin{equation}
\label{funz onda class}
\psi = A(q) e^{i S(q)}.
\end{equation}
The explicit form of the current is
\begin{equation}
\label{j}
j_{0}^{\alpha} = |A|^{2}\nabla^	{\alpha}S.
\end{equation}

Being the conjugated momenta to $q_{\alpha}$ equals to $p_{\alpha}=\nabla_{\alpha}S$, the tangent vector to the classical trajectory can be obtained starting from the variational principal $\delta_{p_{q}}S = 0$, in order to obtain:
\begin{equation}
\label{qpunto}
\dot{q}^{\alpha}=2 N \nabla^{\alpha}S	.
\end{equation}
It is possible to show that, by requiring that the three-dimensional surfaces  $\Sigma_{\alpha}$ on which we defined the probability (\ref{prob3}) are crossed only one time by all the classical trajectories, the sign of the element $\dot{q}^{\alpha}d\Sigma_{\alpha}$ is always the same for any choice of the surface elements $d\Sigma_{\alpha}$. Being the initial sign arbitrary ,we can choose 
\begin{equation}
\label{segno1}
\dot{q}^{\alpha}d\Sigma_{\alpha} > 0.
\end{equation}

The classical current conservation law
\begin{equation}
\label{cons clas vil}
\nabla(A^{2}\nabla S) = 0
\end{equation}
can be recast in a continuity equation form. The first step is the identification of the classical probability distribution $\sigma_{0}=|A|^{2}$. Furthermore, using the relation (\ref{qpunto}) and performing a coordinate transformation for one coordinate of the superspace as $q_{n}=t$, the Eq.(\ref{cons clas vil}) takes the form
\begin{equation}
\label{continuità vilenkin}
\frac{\partial \sigma_{0}}{\partial t} + \partial_{a}\mathcal{J}^{a}=0
\end{equation}
where $\mathcal{J}^{a}=\sigma_{0}\dot{q}^{a}$ and the index a runs from $1$ to $(n-1)$.
From the continuity equation (\ref{continuità vilenkin}) a conserved charge can be identified  integrating both sides over a $d\Sigma_{0}$ volume, where $d\Sigma_{0}$ is the surface element of the subspace defined from the $(n-1)$ remaining classical variables $q^{\alpha}$, and making use of the Gauss Theorem on the current term. This procedure allow to normalize the classical probability distribution as
\begin{equation}
\label{carica classica}
\int \sigma_{0}(q) d\Sigma_{0}=1.
\end{equation} 

The second equation in (\ref{deco}) can be recasts in a Schrodinger-like equation for the quantum subsystem using the relation (\ref{qpunto}):
\begin{equation}
\label{schro 2}
i\frac{\partial\chi}{\partial t} = N H_{\rho}\chi. 
\end{equation}

In order to find the total (classical and quantum) probability distribution  we consider the wave function (\ref{funz onda}) for the current (\ref{corrente1}). This brings to:
\begin{equation}
J = \sigma_{\chi}j_{0}^{\alpha} + \frac{1}{2}|A|^{2}j_{\chi}^{\nu},
\end{equation}
where we defined the quantum current $j_{\chi}^{\nu}$ and probability distribution $\sigma_{\chi}$ as:
\begin{equation}
j_{\chi}^{\nu} = - \frac{i}{2}[\chi^{*}\nabla_{\beta} \chi - \chi\nabla_{\beta} \chi^{*}] \quad , \quad \sigma_{\chi} \equiv |\chi|^{2}.
\end{equation}

From the previous definition of the quantum part of the current and from the Schrodinger equation (\ref{schro 2}), a continuity equation for the quantum probability distribution can be written as
\begin{equation}
\frac{\partial \sigma_{\chi}}{\partial t} + N \nabla_{\nu}j_{\chi}^{\nu}=0,
\end{equation}
To complete the scheme we need to analyze if the total probability distribution is normalizable. It can be written as
\begin{equation}
\sigma(q, \rho) = \sigma_{0}(q)\sigma_{\chi}(q,\rho),
\end{equation} 
where $\sigma_{0}(q)$ is the probability distribution relates to the semiclassical variables. In this case it is possible to show that the surface element of the constant-time surfaces can be written in the form $d\Sigma = d\Sigma_{0} d\Omega_{\rho}$, where $d\Sigma_{0}$ provides the normalization for the classical system:
\begin{equation}
\int \sigma_{0}(q) d\Sigma_{0}=1,
\end{equation}
while $ d\Omega_{\rho}$ gives the normalization for the quantum subsystem:
\begin{equation}
\int \sigma_{\chi}(q,\rho) d\Omega_{\rho}=1.
\end{equation}
As a consequence, the entire probability distribution is normalizable as
\begin{equation}
\label{normalizzazione intera}
\int \sigma_{0}(q)\sigma_{\chi} d\Sigma_{0}d\Omega_{\rho} = 1
\end{equation}

\section{The POLYMER REPRESENTATION OF QUANTUM MECHANICS}
\label{sec:2}
In this Section we briefly summarize the fundamental features of the polymer quantization scheme. In particular, we shall give a general picture of the model, considering the kinematical and dynamical properties.
\subsection{Kinematical properties}
\label{Kin}
The Polymer representation of quantum mechanics is a non-equivalent representation of the usual Schr\"odinger quantum mechanics, based on a different kind of Canonical Commutation Rules (CCR).
The latter represents a very useful tool to study the consequences of the hypothesis for which one or more of the variables of the phase space of the minisuperspace model that is taken into account for the quantization are discretized.

To introduce the basic concepts of the polymer quantum mechanics we start by considering a simple one-dimensional quantum system\cite{corichi},\cite{corichidue}.

Let us consider a set of kets $ |\mu_{i} \rangle$, with $\mu_{i} \in  \mathbb{R}$ and discrete index $i =1,...,N$. The vectors $ |\mu_{i} \rangle$ belong to the Hilbert space $\mathcal{H}_{poly}= L^{2}(\mathbb{R}_{b},d\mu_{H})$\footnote{The set of square-integrable functions defined on the Bohr compactification of the real line $\mathbb{R}_{b}$ with a Haar measure $d\mu_{H}$}. The inner product between two kets is $\langle \nu |\mu \rangle = \delta_{\nu,\mu}$ and the state of the system is described by a generic linear combination of them
\begin{equation}
|\psi\rangle =  \sum\limits_{i=1}^N a_{i}|\mu_{i}\rangle.
\end{equation}
One can identify two fundamental operators in this Hilbert space: a \textit{label operator} $\widehat{\varepsilon}$ and a \textit{shift operator} $\widehat{s}(\lambda)$. They act on the kets as follows
\begin{equation}
\label{label shift}
\widehat{\varepsilon} |\mu \rangle  = \mu |\mu \rangle \quad , \quad \widehat{s}(\lambda) |\mu \rangle = |\mu + \lambda \rangle. \quad , \quad \lambda \in \mathbb{R}^{+}
\end{equation}
where the parameter $\lambda$ is fixed in the quantization procedure but its value can, in principle, span all the real values on the positive axes.
This one-dimensional system is characterized by a the position variable $q$ and the conjugate momenta $p$.
We make the physical choice to assign a discrete characterization to the variable $q$, and to describe the wave function of the system in the so-called $p$-polarization. Consequently, the projection of the states on the pertinent basis vectors is
\begin{equation}
\phi_{\mu}(p) = \langle p | \mu \rangle = e^{-i\mu p}.
\end{equation}
After the introduction of two unitary operators $U(\alpha)=e^{i\alpha \widehat{q}}, V(\beta) = e^{i\beta \widehat{p}} , (\alpha,\beta) \in  \mathbb{R}$ which obey the Weyl Commutation Rules (WCR) $U(\alpha)V(\beta) = e^{i\alpha \beta} V(\beta)U(\alpha)
$\cite{halvorson}, it is possible to show that the label operator represents exactly the position operator, while it is not possible to define a (differential) momentum operator, as a consequence of the discontinuity for $\widehat{s}(\lambda)$ pointed out in Eq.(\ref{label shift}).
\subsection{The dynamical features}
\label{Dyn}
To characterize the dynamical properties of this simple model, it is necessary to investigate the system from the Hamiltonian point of view.
A one-dimensional particle of mass $m$ in a potential $V(q)$ is describing by the Hamiltonian
\begin{equation}
\label{hamlib}
H = \frac{p^{2}}{2m} + V(q).
\end{equation}
Being $q$ a discrete variable, we cannot define, in the $p$-polarization, the operator $\widehat{p}$ as a differential operator. 
The standard procedure to go beyond this problem consists in defining a subspace $ \mathcal{H}_{\gamma_{a}} $ of $ \mathcal{H}_{poly} $. This subspace contains all vectors that live on the lattice of points identified by the lattice spacing $\lambda$
\begin{equation}
\gamma_{\lambda} = \mathcal {f} q \in \mathbb {R} | q = n\lambda, \forall n \in \mathbb {Z} \mathcal {g},
\end{equation}
where $\lambda$ has the dimensions of a \textit{length}.\\
The basis vector takes the form $ | \mu_ {n} \rangle $ (where $ \mu_ {n} = \lambda n $) and the states are defined as a linear combination of them:
\begin{equation}
| \psi \rangle = \sum \limits_{n} b_{n} | \mu _ {n} \rangle.
\end{equation}
The basic realization of the polymer quantization is to approximate the term corresponding to the non-existent operator (this case $\widehat{p}$), and to find for this approximation an appropriate and well-defined quantum operator.
The operator $\widehat{V}$ is exactly the shift operator $\widehat{s}$, in both polarizations. Through this identification, it is possible to exploit the properties of $\widehat{s}$ to write an approximate version of $\widehat{p}$. For $p \ll \frac{\hbar}{\lambda}$, one gets
\begin{equation}
\label{appP}
p \simeq \frac{\hbar}{\lambda}\sin \left( \frac{\lambda p}{\hbar}\right) = \frac{\hbar}{2 i \lambda}\left( e^{i\frac{\lambda p}{\hbar}} - e^{-i\frac{\lambda p}{\hbar}} \right)
\end{equation} 
and then the new version of $\widehat{p}$ is
\begin{equation}
\widehat{p}_{\lambda} |\mu_{n} \rangle = \frac{i \hbar}{2\lambda}\left( |\mu_{n-1}\rangle - |\mu_{n+1} \rangle \right).
\end{equation}
One can define an approximate version of $\widehat{p}^{2}$. For $p \ll \frac{\hbar}{\lambda}$, one gets
\begin{equation}
\label{appP2}
p^{2} \simeq \frac{2 \hbar^{2}}{\lambda^{2}}\left[1-\cos\left( \frac{\lambda p}{\hbar} \right) \right] = \frac{2 \hbar^{2}}{\lambda^2} \left[1 - e^{i\frac{\lambda p}{\hbar}} - e^{-i\frac{\lambda p}{\hbar}} \right]
\end{equation}
and then the new version of $\widehat{p}^{2}$ is
\begin{equation}
\widehat{p}_{\lambda}^{2} |\mu_{n} \rangle =  \frac{\hbar^{2}}{\lambda^2} \left[ 2|\mu_{n}\rangle - |\mu_{n+1}\rangle - |\mu_{n-1}\rangle \right].
\end{equation}
Remembering that $\widehat{q}$ is a well-defined operator as in the canonical way, the approximate version of the starting Hamiltonian (\ref{hamlib}) is
\begin{equation}
\label{Hpoly}
\widehat{H}_{\lambda} = \frac{1}{2m}\widehat{p}_{\lambda}^{2} + V(\widehat{q}).
\end{equation}
The Hamiltonian operator $\widehat{H}_{\lambda}$ is a well-defined and symmetric operator belonging to $ \mathcal{H}_{\gamma_{\lambda}} $.

\section{Bianchi I model in presence of a stiff matter contribution}
\label{sec:3}
In this Section we introduce a simple and instructive model for which it is possible to individuate a separation in the configuration space between classical and quantum variables. Let us consider a universe described by a Bianchi I model filled with a stiff matter contribution.
The description of the model will be done with respect to the Misner-like variables $\{a,\beta_{\pm}\}$\footnote{The original isotropic Misner variable $\alpha$ is just $\alpha = \ln a$. }, where $a$ expresses the isotropic volume of the universe (the initial singularity is reached for $a\rightarrow0$) while $\beta_{\pm}$ accounts for the anisotropies of this model. Although the stiff matter term does not own the same dynamical proprieties, its equation of state mimic the one of a free massless scalar field. For this reason, in presence of a stiff matter contribution, recalling the form of the superHamiltonian in presence of a massless scalar field\cite{scalarfield},\cite{berger}, the scalar constraint of the Bianchi I model takes the form\footnote{We use the $(-,+,+,+)$ signature of the metric, the unit system with $c=1$ and we explicit the Einstein constant as $k=8\pi G = \frac{8 \pi l_{p}^{2}}{\hbar}$ where $l_{p}$ is the Planck length.}
\begin{equation}
\label{hamiltoniana}
\mathcal{H} = \frac{l_{p}^{2}}{24 \pi \hbar}\left[ -\frac{p_{a}^{2}}{a} + \frac{p_{+}^{2} + p_{-}^{2}}{a^{3}}\right] + \frac{8 \pi^{2} \mu^{2}}{\hbar a^{3}} = 0
\end{equation}
where $ \{ p_{a},p_{+},p_{-} \} $ are the conjugated momenta related to the Misner-like variables and the constant $\mu$ represents the stiff matter contribution.
The canonical quantization of the model will be done at first in the $a$-polarization, following the prescription of the Section \ref{sec:1}, and then in the $p_{a}$-polarization. In both cases the quantization of the anisotropies will be in the position polarization. Through the realization of such a comparison we shall show how the semiclassical and quantum solutions obtained will be equivalent in both cases. This result will be very useful in respect of the implementation of the polymer paradigm.
\subsection{$a$-polarization}
\label{3.1}
Here we perform a canonical quantization imposing that the physical states $\psi$ being annihilated by the operator $\mathcal{H}$, \textit{i.e.} the quantum version of the superHamiltonian constraint (\ref{hamiltoniana}).
If we choose to describe all the configuration variables in the position polarization, this means that \{$\widehat{a},\widehat{\beta}_{+},\widehat{\beta}_{-}$ \}  act as a multiplicative operators and \{$\widehat{p}_{a},\widehat{p}_{+},\widehat{p}_{-}$ \} as a derivative operators in this way:
\begin{equation}
\widehat{p}_{a} = -i\hbar \frac{\partial}{\partial a}=-i\hbar\partial_{a} \quad , \quad \widehat{p}_{\pm} = -i\hbar \frac{\partial}{\partial \beta_{\pm}}=-i\hbar\partial_{\pm}.
\end{equation}
Therefore, the WDW equation for the superHamiltonian (\ref{hamiltoniana}) can be written as
\begin{equation}
\label{WDW 2}
\left[\hbar^{2} a^{2} \partial_{a}^{2} - \hbar^{2}\left(\partial_{+}^{2} + \partial_{-}^{2}\right) +\frac{3 (4 \pi)^{3}\mu^{2}}{l_{p}^{2}} \right]\psi(a,\beta_{\pm})=0.
\end{equation} 
Starting from the equation (\ref{WDW 2}), it is possible to individuate a corresponding current of this form 
\begin{displaymath}
J^{\mu} =
\left[ \begin{array}{ccc}
J^{a} \\
J^{+} \\
J_{-}
\end{array} \right] = -\frac{i}{2}\hbar^{2}\left[ \begin{array}{ccc}
a^{2} ( \psi^{*} \partial_{a}\psi - \psi \partial_{a}\psi^{*}) \\
-( \psi^{*} \partial_{+}\psi - \psi \partial_{+}\psi^{*})  \\
-( \psi^{*} \partial_{-}\psi - \psi \partial_{-}\psi^{*})
\end{array} \right],
\label{corr stan a}
\end{displaymath}
for which the conservation law $\nabla_{\mu}J^{\mu} = 0$ is valid.
Here  $\nabla_{i}$ is the covariant derivative built with the superspace metric $g^{\mu \nu}=diag(\hbar^{2} a^{2},-\hbar^{2},-\hbar^{2})$ and its action on a generic vector $v^{\nu}$ is
\begin{equation}
\label{azione derivata covariante}
\nabla_{\mu}v^{\nu} = \partial_{\mu}v^{\nu} + \Gamma^{\nu}_{\mu \rho}v^{\rho}.
\end{equation}
For our superspace metric $g_{\mu \nu}$, the only non-vanishing Christoffel symbol is $\Gamma^{a}_{aa} = -\frac{2}{a}$. Therefore, the conservation of the current in Eq.(\ref{WDW 2}) takes the form:
\begin{equation}
\label{conservazione esplicita a}
\nabla_{\mu}J^{\mu}= \partial_{\mu}J^{\mu} + \Gamma^{\mu}_{\mu \rho}j^{\rho} = \left( \partial_{a} - \frac{2}{a}\right)J^{a} + \partial_{+}J^{+} + \partial_{-}J^{-}=0.
\end{equation}
 
Following the Vilenkin interpretation of the wave function discussed in the Section \ref{sec:1}, as it is natural we choose to assign the character of \textit{semiclassical} to the isotropic variable $a$, while the anisotropies $\{\beta_{+},\beta_{-}\}$ characterize the \textit{quantum} subsytem. With this prescription we choose the wave function of the Universe as:
\begin{equation}
\label{funz a}
\psi(a,\beta_{\pm}) = \chi(a)\varphi(a,\beta_{\pm})= A(a)e^{\frac{i}{\hbar}S(a)}\varphi(a,\beta_{\pm}).
\end{equation}
Considering the previous wave function inside the WDW equation (\ref{WDW 2}) we obtain, at the lowest order in  $\hbar$, the Hamilton-Jacobi equation:
\begin{equation}
\label{HJ 2}
-a^{2}(S')^{2} + \frac{3 (4 \pi)^{3}\mu^{2}}{l_{p}^{2}} = 0,
\end{equation}
where $(\bullet )' \equiv \frac{\partial}{\partial a}$.
In the equation (\ref{HJ 2}) does not appear the anisotropies or their conjugate momenta; in fact, the lowest order in the WKB performed respect to the $\hbar$ parameter takes into account the semiclassical behavior of the whole system, and this regard only the isotropic variable. Furthermore, making a comparison between the Eq.(\ref{HJ 2}) and the superHamiltonian constraint (\ref{hamiltoniana}) when the anisotropies are ``frozen''\footnote{In this case we mean that the term $\frac{p_{+}^{2} + p_{-}^{2}}{a^{3}}$ is negelcted.} , we can establish the connection $S' = p_{a}$ and rewrite the Eq.(\ref{HJ 2}) in this way:
\begin{equation}
\label{impulso1}
p_{a}^{2} = \frac{3 (4\pi)^{3}\mu^{2}}{l_{p}^{2}a^{2}}.
\end{equation}
It is possible to obtain the explicit solution for $a=a(t)$ making use of the Eq.(\ref{impulso1}) with the Hamiltonian Equation
\begin{equation}
\label{Hamilton1}
\frac{d a}{d t} = \frac{\partial \mathcal{H}}{\partial p_{a}} = -\frac{l_{p}^{2}}{12 \pi \hbar }\frac{p_{a}}{a}
\end{equation}
in order to achieve\footnote{We choose the integration's constant in such a way that $a(0)=0$.}
\begin{equation}
\label{diff1}
\frac{d a}{d t} = \sqrt{\frac{4 \pi}{3}}\frac{l_{p} \mu}{\hbar a^{2}} \Longrightarrow a(t)=\left(\frac{2 \sqrt{3 \pi } l_{p} \mu}{h} t\right)^\frac{1}{3}.
\end{equation}
As we can see from the Eq.(\ref{diff1}), the zero value of the isotropic variable, reached for $t=0$, exhibits the singular behavior of the model.

The next order in the WKB expansion gives us the following equation
\begin{equation}
\label{primo ordine}
i a^{2} \frac{1}{A}(A^{2}S')' + 2 i a^{2} A S' \varphi' - A \left(\frac{\partial^{2}}{\partial \beta_{+}^{2}} + \frac{\partial^{2}}{\partial \beta_{-}^{2}}\right)\varphi = 0.
\end{equation}
As in the Section \ref{sec:1}, we can decoupled the above equation making an adiabatic approximation. Naturally, we require that the $a$-evolution is mainly contained in the amplitude $A$, while the isotropic variation of the quantum part $\varphi$ is negligible. This is express by the condition
\begin{equation}
\label{app adia a}
|\partial_{a} \chi(a)| \gg |\partial_{a} \varphi(a,\beta_{\pm})|.
\end{equation}
Considering the condition 
(\ref{app adia a}) in the Eq.(\ref{primo ordine}), we obtain:
\begin{equation}
\label{deco a}
 \frac{a^{2}}{A}(A^{2}S')' = 0 \quad , \quad  2 i a^{2}S' \varphi' -   \left(\partial_{+}^{2} + \partial_{-}^{2}\right)\varphi = 0
\end{equation}

Looking at the first equation in (\ref{deco a}), we can see that it corresponds to the conservation of the current $\nabla_{a}J^{a} = 0$ when we take into account just the semiclassical version of the wave function (\ref{funz a}):
\begin{equation}
\psi(a) = A(a)e^{\frac{i}{\hbar}S(a)}.
\end{equation}
The explicit form of the current is 
\begin{equation}
\label{corr classica}
J^{a} =\hbar a^{2}A^{2}S'.
\end{equation}

The second equation in (\ref{deco a}) provides the evolution of the quantum subsystem:
\begin{equation}
\label{quantistica a}
2 i a^{2} S'\partial_{a}\varphi= \left(\partial_{+}^{2} + \partial_{-}^{2}\right)\varphi
\end{equation}
It is important to underline that, in analogy with the Vilenkin approach, the Eq.(\ref{quantistica a}) is consistent requiring that $\left(\partial_{+}^{2} + \partial_{-}^{2}\right)\varphi= O(\hbar)$. It is possible to write a Schrodinger-like equation for the quantum wave function $\varphi$ using the relation $ \frac{\partial \varphi}{\partial a} = \frac{\partial \varphi}{\partial t}\frac{\partial t}{\partial a} $ and the Eqs.(\ref{impulso1}),(\ref{diff1}). This way we have:
\begin{equation}
\label{schro 1}
i \left(\frac{24 \pi \hbar}{l_{p}^{2}}\right)a^{3}\frac{\partial \varphi (t, \beta_{\pm})}{\partial t}= \left(\partial_{+}^{2} + \partial_{-}^{2}\right)\varphi(t, \beta_{\pm}),
\end{equation}
and with the introduction of the time-like variable $\tau$ for which $\frac{\partial }{\partial \tau} = \frac{24 \pi}{l_{p}^{2}}a^{3}\frac{\partial }{\partial t}$ we can finally write:
\begin{equation}
\label{eq onde 1}
i\hbar\frac{\partial \varphi (\tau, \beta_{\pm})}{\partial \tau} = \left(\partial_{+}^{2} + \partial_{-}^{2}\right)\varphi(\tau, \beta_{\pm}).
\end{equation}
The Eq.(\ref{eq onde 1}) resembles a plane wave equation which solution is of the form
\begin{equation}
\label{onda piana}
\varphi(\tau,\beta_{\pm}) = e^{ \frac{iE \tau}{\hbar}}e^{\frac{i k_{+}\beta_{+}}{\hbar}}e^{\frac{i k_{-}\beta_{-}}{\hbar}},
\end{equation}
with $E=\frac{(k_{+}^{2} + k_{-}^{2})}{\hbar^{2}}$. We can make explicit the dependence $\tau(t)$ by solving the integral:
\begin{equation}
\label{tau a}
\tau(t)=\int \frac{l_{p}^{2}}{24\pi a(t)^{3}}dt = \frac{l_{p}\hbar}{48\pi \sqrt{3 \pi}\mu}\ln \frac{t}{t^{*}},
\end{equation}
where $t^{*}$ is the integration constant. 
To conclude, the quantum part of the wave function takes the form:
\begin{equation}
\label{funz onda qua a}
\varphi(t,\beta_{\pm}) = Ce^{ i\frac{l_{p}}{48\pi \sqrt{3 \pi}\hbar^{2}}(k_{+}^{2} + k_{-}^{2})\ln \frac{t}{t^{*}}}e^{\frac{i k_{+}\beta_{+}}{\hbar}}e^{\frac{i k_{-}\beta_{-}}{\hbar}},
\end{equation}
where the $\{k_{+},k_{-} \}$ are the quantum numbers associated to the anisotropies, for which is valid the dispersion relation $k_{\pm} = p_{\pm}$.

\subsection{$p_{a}$-polarization}
\label{3.2}
This subsection is devoted to the implementation of the quantum  model in the $p_{a}$-polarization. The action of the operators \{$\widehat{p}_{a},\widehat{\beta}_{+},\widehat{\beta}_{-}$ \}  is multiplicative, while \{$\widehat{a},\widehat{p}_{+},\widehat{p}_{-}$ \} act as a derivative operators:
\begin{equation}
\label{azione operatori p}
\widehat{a} = i\hbar \frac{\partial}{\partial p_{a}} = i\hbar\partial_{p_{a}} \quad , \quad \widehat{p}_{\pm} = -i\hbar \frac{\partial}{\partial \beta_{\pm}}=-i\hbar \partial_{\pm}.
\end{equation}
The quantum counterpart of the superHamiltonian constraint (\ref{hamiltoniana}), or in other words the WDW equation, in this polarization is:
\begin{equation}
\label{WDW 3}
\left[\hbar^{2} p_{a}^{2} \partial_{p_{a}}^{2} - \hbar^{2}\left(\partial_{+}^{2} + \partial_{-}^{2}\right) +\frac{3 (4 \pi)^{3}\mu^{2}}{l_{p}^{2}} \right]\psi(p_{a},\beta_{\pm})=0
\end{equation} 

Also in this case,  with the difference that the role of $a$ is taken by the conjugated momenta $p_{a}$, choosing the normal operator-ordering for the isotropic part of the WDW equation from the Eq.(\ref{WDW 3}) we can build a current term this way
\begin{displaymath}
\label{corr stan p}
J^{\mu} =
\left[ \begin{array}{ccc}
J^{p_{a}} \\
J^{+} \\
J_{-}
\end{array} \right] = -\frac{i}{2}\hbar^{2} \left[ \begin{array}{ccc}
 p_{a}^{2} ( \psi^{*} \partial_{p_{a}}\psi - \psi \partial_{p_{a}}\psi^{*}) \\
-( \psi^{*} \partial_{+}\psi - \psi \partial_{+}\psi^{*})  \\
-( \psi^{*} \partial_{-}\psi - \psi \partial_{-}\psi^{*})
\end{array} \right],
\end{displaymath}
that respect the conservation law $\nabla_{i}J^{i}=0$.

The operator $\nabla_{i}$, in this polarization, is the covariant derivative built with the superspace metric $h^{ij}=diag(\hbar^{2}p_{a}^{2},-\hbar^{2},-\hbar^{2})$. 
Again, there is just one non-vanishing Christoffel symbol $\Gamma^{p_{a}}_{p_{a}p_{a}} = -\frac{2}{p_{a}}$ and the conservation of the current in Eq.(\ref{WDW 3}) takes the form:
\begin{equation}
\label{conservazione esplicita p}
\nabla_{i}J^{i}= \partial_{i}J^{i} + \Gamma^{i}_{ik}j^{k} = \left( \partial_{p_{a}} - \frac{2}{p_{a}}\right)J^{p_{a}} + \partial_{+}J^{+} + \partial_{-}J^{-}=0.
\end{equation}

As in the previous subsection, we choose to assign the role of the quantum subsystem to the anisotropies while the semiclassical variable is  represented by $p_{a}$.
A straightforward version of the wave function of the Universe is:
\begin{equation}
\label{funz p}
\psi(p_{a},\beta_{\pm}) = \chi(p_{a})\varphi(p_{a},\beta_{\pm}) = A(p_{a})e^{\frac{i}{\hbar}S(p_{a})}\varphi(p_{a},\beta_{\pm}).
\end{equation}
Considering the above shape for the wave function in the WDW equation leads, to the lowest order in $\hbar$, to the Hamilton Jacobi equation:
\begin{equation}
\label{HJ 3}
-p_{a}^{2}(\dot{S})^{2} + \frac{3 (4 \pi)^{3}\mu^{2}}{l_{p}^{2}} = 0,
\end{equation}
where $\dot{(\bullet)} \equiv \frac{\partial}{\partial p_{a}}$. This time, a comparison between the Eq.(\ref{HJ 3}) and the part related to the semiclassical variable of the superhamiltonian constraint establish the equality $\dot{S} = a$ and the H-J equation can be written as: 
\begin{equation}
\label{impulso2}
p_{a}^{2} = \frac{3 (4\pi)^{3}\mu^{2}}{l_{p}^{2}a^{2}},
\end{equation}
exactly the same relation obtained in the Eq.(\ref{impulso1}) for the $a$-polarization case. As a consequence, also the evolution of $a(t)$ is the same in the Eq.(\ref{diff1}).

If we consider the successive order in the WKB expansion we obtain the equation
\begin{equation}
\label{primo ordine p}
i p_{a}^{2}\frac{1}{A}(A^{2}\dot{S})^{\dot{}} + 2 i p_{a}^{2} A \dot{S} \dot{\varphi} - A \left(\partial_{+}^{2} + \partial_{-}^{2}\right)\varphi = 0.
\end{equation}

Via the same consideration of the previous subsection, we can obtain from the Eq.(\ref{primo ordine p}) a pair of equations using the adiabatic approximation that, in the $p_{a}$-polarization, is expresses by the condition:
\begin{equation}
\label{app adia p}
|\partial_{p_{a}} \chi(p_{a})| \gg |\partial_{p_{a}} \varphi(p_{a},\beta_{\pm})|.
\end{equation}
This way we obtain:
\begin{equation}
\label{deco p}
 \frac{p_{a}^{2}}{A}(A^{2}\dot{S})^{\dot{}} = 0 \quad , \quad  2 i p_{a}^{2} A \dot{S} \dot{\varphi} - A \left(\partial_{+}^{2} + \partial_{-}^{2}\right)\varphi = 0.
\end{equation}

The first equation in (\ref{deco p}) corresponds to the conservation of the current $\nabla_{p_{a}}J^{p_{a}}=0$  when the semiclassical version of the wave function (\ref{funz p}) is considered:
\begin{equation}
\psi(a) = A(a)e^{\frac{i}{\hbar}S(a)}.
\end{equation}
The current obtained in this way has the form
\begin{equation}
\label{corr classica}
J^{a} =\hbar p_{a}^{2}A^{2}\dot{S}.
\end{equation}

The description of the quantum subsytem is contained in the second equation in (\ref{deco p}):
\begin{equation}
\label{quntistica p}
 2 i p_{a}^{2} \dot{S} \dot{\varphi} = \left(\partial_{+}^{2} + \partial_{-}^{2}\right)\varphi.
\end{equation}
An explicit time dependence for the $\varphi$ can be introduce using the relation $ \frac{\partial \varphi}{\partial p_{a}} = \frac{\partial \varphi}{\partial t}\frac{\partial t}{\partial p_{a}} = \frac{\partial \varphi}{\partial t}\frac{1}{\dot{p_{a}}}$. The quantity $\dot{p_{a}}$ can be evaluated differentiating the relation (\ref{impulso2}) and using the Eq.(\ref{diff1}) it is possible to recast the  Eq.(\ref{quntistica p}) in such a way:
\begin{equation}
\label{schro 2}
i \left(\frac{24 \pi \hbar}{l_{p}^{2}}\right)a^{3}\frac{\partial \varphi (t, \beta_{\pm})}{\partial t} =  \left(\partial_{+}^{2} + \partial_{-}^{2}\right)\varphi(t, \beta_{\pm}).
\end{equation}
A comparison between the last equation and the Eq.(\ref{schro 1}) shows that the two differential equation are the same. Furthermore we can proceed in the same way as in the previous subsection and conclude that the solution for the wave function related to the quantum subsystem is
\begin{equation}
\label{funz onda qua p}
\varphi(t,\beta_{\pm}) = Ce^{ i\frac{l_{p}}{48\pi \sqrt{3 \pi}\hbar^{2}}(k_{+}^{2} + k_{-}^{2})\ln \frac{t}{t^{*}} }e^{\frac{i k_{+}\beta_{+}}{\hbar}}e^{\frac{i k_{-}\beta_{-}}{\hbar}}.
\end{equation}

As we expected, also in the  in the implementation of the Vilenkin approach there are no differences, both from a semiclassical and the quantum point of view,  when we study the problem in the position polarization respect to the momentum polarization. This aspect will be crucial in the next Section, when the discrete nature of the isotropic variable shall be taken into account in the context of the Polymer Quantum Mechanics.

Let us now say something about the application of the Vilenkin steps for the individuation of a conserved probability distribution. In the general scheme illustrated in Section \ref{sec:1}, the coordinated transformation $h_{n}=t$ for one of the classical configuration variable allowed to rewrite the conservation law (\ref{cons clas vil}) in the continuity equation form (\ref{continuità vilenkin}) and to individuate a normalizable probability distribution (\ref{carica classica}). The crucial point in this procedure is the fact that the classical configuration space contains more than one variable, in order to define an orthogonal $\Sigma_{0}$ surfaces over which integrate. In our model, equally in both polarization, the classical configuration space has dimension one (we have just one variable: $a$ or $p_{a}$). This imply that the orthogonal space over which to evaluate the probability distribution has dimension zero and therefore the Vilenkin procedure cannot be replicated.

Regarding the quantum sub-system, starting from the quantum part of the wave function $\varphi$ in the Eq.(\ref{funz onda qua a}), a probability distribution for the quantum variables is defined as $\rho_{\varphi}=|\varphi|^{2}$. This way, the leading terms of the components of the current (\ref{WDW 2}), considering the entire wave function (\ref{funz onda qua a}), assume the form:
\begin{equation}
\label{clas a base a }
J^{a} =\hbar a^{2}A^{2}S'\rho_{\varphi},
\end{equation}
\begin{equation}
\label{clas beta base a}
J^{\pm} = -\frac{\hbar^{2}A^{2}}{2}( \varphi^{*}\partial_{\pm}\varphi - \varphi\partial_{\pm}\varphi^{*} ) \equiv \frac{A^{2}}{2}J^{\pm}_{\varphi},
\end{equation}
and the conservation law $\nabla_{\mu}J^{\mu}=0$ can be recast as
\begin{equation}
\label{eq cont 1}
2\hbar a^{2}S'\frac{d\varphi}{d a} + \partial_{i}J^{i}_{\varphi}=0.
\end{equation}
We provides the calculus in the $a$-polarization, but the conclusions will be the same also in the other one.
In the above rewrite of the conservation of the current the index $i=\{+,-\}$ and we used the first relation in the Eq.(\ref{deco a}). 
Then, an explicit presence of the variable $t$ can be include through the relation $\frac{\partial \varphi}{\partial a} = \frac{\partial \varphi}{\partial t}\frac{\partial t}{\partial a} $ and making use of the Eqs.(\ref{HJ 2}),(\ref{diff1}), in order to obtain a \textit{continuity equation}:
\begin{equation}
\label{continuita a}
\frac{d \rho_{\varphi}}{dt} = -\frac{l_{p}^{2}}{24 \hbar^{2}\pi a^{3}(t)}\partial_{i}J^{i}_{\varphi}.
\end{equation}
Integrating the both sides of the equation over a ${\beta_{+},\beta_{-}}$ volume we have that the right side can be rewritten using the Gauss Theorem:
\begin{equation}
\label{lato destro}
\frac{l_{p}^{2}}{24 \hbar^{2}\pi a^{3}(t)}\int \int d\beta_{+}d\beta_{-}\partial_{i}J^{i}_{\varphi} = \frac{l_{p}^{2}}{24 \hbar^{2}\pi a^{3}(t)}\int_{\partial\beta}d\sigma J^{i}_{\varphi}
\end{equation}
Making the hypothesis that all the system is contained in the surface $\partial\beta$, the term in the Eq.(\ref{lato destro}) vanishes, being evaluated over the surfaces. Therefore, what remains in the integration of the Eq.(\ref{continuita a}) is
\begin{equation}
\frac{d}{dt}\int \int d\beta_{+}d\beta_{-}\rho_{\phi} = 0,
\end{equation}
which means that the integral is conserved and can be normalized as
\begin{equation}
\label{normalizzazione stan}
\int \int d\beta_{+}d\beta_{-}\rho_{\varphi} = 1
\end{equation}
All the previous steps can be replicated in the $p_{a}$-polarization simply taking in consideration the conservation of the current (\ref{WDW 3}) with the wave function (\ref{funz onda qua p}) and through the relation $\frac{\partial \varphi}{\partial p_{a}} = \frac{\partial \varphi}{\partial t}\frac{\partial t}{\partial p_{a}}$. This leads exactly to the continuity equation (\ref{continuita a}) and to the same final considerations. 

To conclude this Section we underline how, for our specific model, it was not possible to build an entire conserved probability distribution as in the Eq.(\ref{normalizzazione intera}). Nevertheless, our results are not so far from the Vilenkin conclusions. Indeed, we were able to obtain a quantum normalizable probability distribution (\ref{normalizzazione stan}) coupled with a classical conserved quantity (\ref{deco a}),(or the Eq.(\ref{deco p}) in the $p_{a}$-polarization).
\subsection{Adiabatic Approximation}
We go ahead in the analysis of the model performing a test about the validity of the adiabatic approximation in the two polarization cases (\ref{app adia a}),(\ref{app adia p}).

First of all, considering the conservation of the classical current in the Eq.(\ref{deco a}) and the H-J equation (\ref{HJ 2}), we can argue that amplitude of the wave function of the Universe evolves as $A(a)\propto a^{\frac{1}{2}}$ and therefore $|\partial_{a}A(a)|\propto  a^{-\frac{1}{2}}$. Furthermore, using the Eq.(\ref{diff1}), for the derivative of the quantum part of the wave function (\ref{funz onda qua a}) the behavior is $|\partial_{a}\varphi| \propto a^{-1}$. Given the identification  $S' = p \propto \frac{1}{a}$, for what concerns the $a$-polarization case we can then conclude that the adiabatic approximation (\ref{app adia a}) is valid for an initial condition of the Universe in which the isotropic variable, or in other words the volume of the Universe, assumes not too small values.

Repeating the same steps in the $p$-polarization, from the conservation of the classical current (\ref{deco p}) and the H-J equation (\ref{HJ 3}) we can achieve that $A(p_{a})\propto p_{a}^{\frac{1}{2}}$ and consequently $|\partial_{p_{a}}A(p_{a})|\propto  p_{a}^{-\frac{1}{2}}$. Using the H-J equation and the Eq.(\ref{diff1}), the derivative of the wave function $\varphi$ behaves as $|\partial_{p_{a}}\varphi| \propto p_{a}^{-1}$. Taking into account the founded trends, we can conclude, recalling $\dot{S} = a \propto \frac{1}{p}$, that the adiabatic approximation (\ref{app adia p}), is valid in the $p$-polarization when the conjugated momenta to the isotropic variables starts its evolution towards the singularity assuming large values.

\subsection{Expectation values of the anisotropies: the Ehrenfest theorem}
To conclude this Section let us analyze the behavior of the quantum variables: the anisotropies. To this aim, let us introduce a useful theorem to study the evolution of a quantum operator: the \textit{Ehrenfest Theorem}. Let $|\varphi>|$ the state of the quantum subsytem built starting by the wave function (\ref{funz onda qua a}). The expectation value of the quantum operators $\{ \widehat{\beta}_{+},\widehat{\beta}_{-} \}$ corresponds to:
\begin{equation}
\label{valore medio}
<\widehat{\beta}_{\pm}> = <\varphi | \widehat{\beta}_{\pm} | \varphi>,
\end{equation}
where $| \varphi>$ is a ket built from the quantum part of the wave function (\ref{funz onda qua a}).
The time derivative of the expectation value of a time independent operator $A$ is given by\footnote{This general equality is due to a Werner-Heisenberg theorem}
\begin{equation}
\frac{d}{d t}<A> = \frac{1}{i \hbar}<[A,H]>,
\end{equation}
where $H$ is the Hamiltonian of the system.
The Ehrenfest Theorem concerns the application of the above results to the one-dimensional systems. Therefore, remembering that the only non vanishing position-momentum commutators are $[\beta_{+},p_{+}]=[\beta_{-},p_{-}]=i\hbar$ , we can apply it to the anisotropies in order to obtain
\begin{multline}
\label{ehren stan posi}
\frac{d<\widehat{\beta}_{\pm}>}{dt} = \frac{1}{i \hbar}<[\beta_{\pm},\mathcal{H}]> = \\ =\frac{l_{p}^{2}}{24 i \pi \hbar^{2} a^{3}}<[\beta_{\pm},p_{\pm}^{2}]> = \frac{l_{p}^{2}}{12 \pi \hbar a^{3}}<p_{\pm}>,
\end{multline}
where we used the commutation rule $[A,BC] = [A,B]C + B[A,C]$.
In the previous relation the term $\frac{1}{a^{3}}$ was brought out to the expectation values because is evaluated over the quantum states $|\varphi(a,\beta_{\pm})>$ and the isotropic variable represent for them a fixed orbit over which the dynamics of the anysotropies occurs.
Applying the Ehrenfest theorem to $p_{\pm}$ it is possible to show that its expectation value is a constant of motion :
\begin{equation}
\label{ehren stan imp}
\frac{d<\widehat{p_{\pm}}>}{dt} = \frac{1}{i \hbar}<[p_{\pm},\mathcal{H}]> = 0 \rightarrow <\widehat{p_{\pm}}> = const.
\end{equation}
Using the above result in the Eq.(\ref{ehren stan posi}) with the time-evolution for the isotropic variable in the Eq.(\ref{diff1}) we arrive at the differential equation:
\begin{equation}
\label{eq diff beta stan}
\frac{d<\widehat{\beta}_{\pm}>}{dt} = \frac{l_{p} p_{\pm}}{24 \sqrt{3\pi} \pi \mu}\frac{1}{t}
\end{equation}
whose solution is
\begin{equation}
\label{sol beta stan}
<\widehat{\beta}_{\pm}>_{t} = \frac{l_{p} p_{\pm}}{24 \sqrt{3\pi} \pi \mu}\ln \frac{t}{t^{*}},
\end{equation}
where $t^{*}$ is an integration constant. 
\begin{figure}[h!]
\includegraphics[scale=.67]{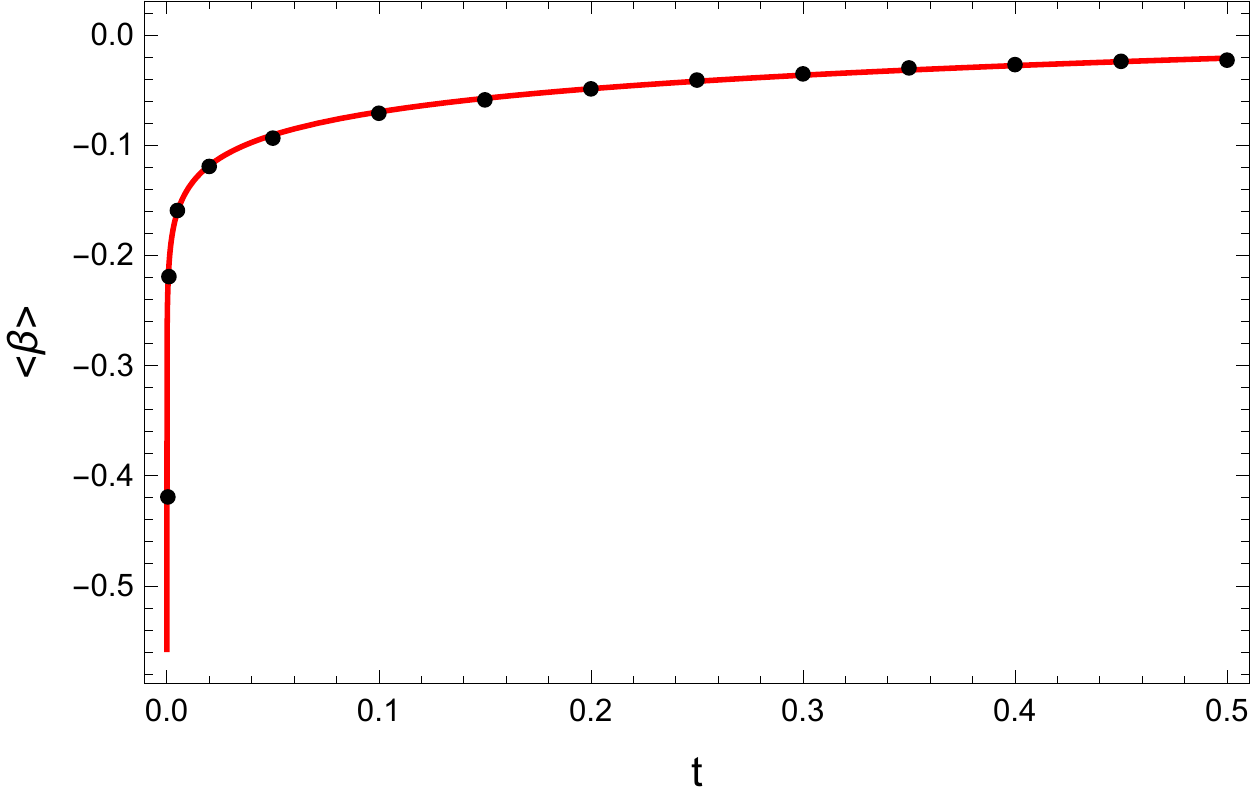}
\caption{The black points represent the position of the maximum of the wave packet $|\Psi_{k_{\pm}^{*}}(t,\beta_{\pm})|$
evaluated via numerical integration for the following choice of
the integration parameters:$C=1,k_{+}^{*} = k_{-}^{*} = 1,\sigma_{+} =\sigma{-} = 0.03$. The continuous red line represents the trajectory evaluated with the same parameters from the Ehrenfest theorem.}
\label{fig1}
\end{figure}

From the Eq.(\ref{sol beta stan}) we can see that the anisotropies become important in magnitude near the initial singularity and they diverge for $t=0$, as we expected from any classical anisotropic model.
From the study of the standard deviation it is possible to establish effectively if the trajectory obtained in the Eq.(\ref{sol beta stan poly}) differs not to much from the classical trajectory.

The application of the Ehrenfest Theorem to the operator $<\widehat{\beta}^{2}>$ brings to:
\begin{multline}
\label{ehren stan posi quad}
\frac{d<\widehat{\beta}_{\pm}^{2}>}{dt} = \frac{1}{i \hbar}<[\beta_{\pm}^{2},\mathcal{H}]> =\frac{l_{p}^{2}<[\beta_{\pm}^{2},p_{\pm}^{2}]>}{24 i \pi \hbar^{2} a^{3}} = \\ = \frac{l_{p}^{2}}{12 \pi \hbar a^{3}}<\beta_{\pm}p_{\pm} + p_{\pm}\beta_{\pm}>,
\end{multline}
where we used the commutation rule $[AB,CD] = AC[B,D] + A[B,C]D+C[A,D]B+[A,C]DB$.
The quantity $<\beta_{\pm}p_{\pm} + p_{\pm}\beta_{\pm}>$ can be evaluated applying again the Ehrenfest theorem:
\begin{multline}
\frac{d <\beta_{\pm}p_{\pm} + p_{\pm}\beta_{\pm}>}{dt} = \\ = \frac{l_{p}^{2}<[\beta_{\pm}p_{\pm} + p_{\pm}\beta_{\pm},p_{\pm}^{2}]>}{24 i \pi \hbar^{2} a^{3}} =\frac{l_{p}^{2}p_{\pm}^{2}}{6 \pi \hbar a^{3}}.
\end{multline} 
The above differential equation can be solved using the time-dependence (\ref{diff1}) in order to obtain:
\begin{equation}
\label{sol intermedia}
<\beta_{\pm}p_{\pm} + p_{\pm}\beta_{\pm}> = \frac{l_{p} p_{\pm}^{2}}{12\pi\sqrt{3 \pi}\mu}(\ln t + K),
\end{equation}
where $K$ is an integration constant.
Inserting the latter relation in the Eq.(\ref{ehren stan posi quad}) we obtain the solution
\begin{equation}
\label{sol beta quadro stan}
<\widehat{\beta}_{\pm}^{2}> = \frac{l_{p}^{2}p_{\pm}^{2}}{(12\pi)^{3}\mu ^{2}}\left[ \ln^{2}t + 2K\log \frac{t}{t^{*}} - \ln^{2}t^{*} \right].
\end{equation}
Finally, it is possible to write the standard deviation for the operator $<\widehat{\beta}_{\pm}>$ as
\begin{multline}
\label{dev stan}
\sigma_{\beta} = \sqrt{<\widehat{\beta}_{\pm}^{2}> -<\widehat{\beta}_{\pm}>^{2} } = \\ = \frac{l_{p} p_{\pm}}{24 \sqrt{3\pi} \pi \mu}\sqrt{2\left(-\ln^{2}\tau + K \ln \frac{t}{\tau} + \ln t\ln \tau\right)}
\end{multline}
The presence of the square root in the standard deviation (\ref{dev stan}) force to impose, in order to maintain a physical meaning for this quantity, particular values to the integration constant $K$. From the Eq.(\ref{dev stan}) we see  that also the standard deviation shows a divergent nature in proximity of the singularity, as it is clear performing the limit $t\rightarrow 0$. However, An estimate of how the expectation value (\ref{sol beta stan}) differs from the classical trajectory can be evaluated from the ratio
\begin{equation}
\label{ratio stan}
\frac{\sigma_{\beta}}{<\widehat{\beta}_{\pm}>} = \frac{\sqrt{2\left( -\ln^{2}\tau + K \ln \frac{t}{\tau} +\ln t\ln \tau\right)}}{\ln \frac{t}{\tau}}
\end{equation}
\begin{figure}[h!]
\includegraphics[scale=.67]{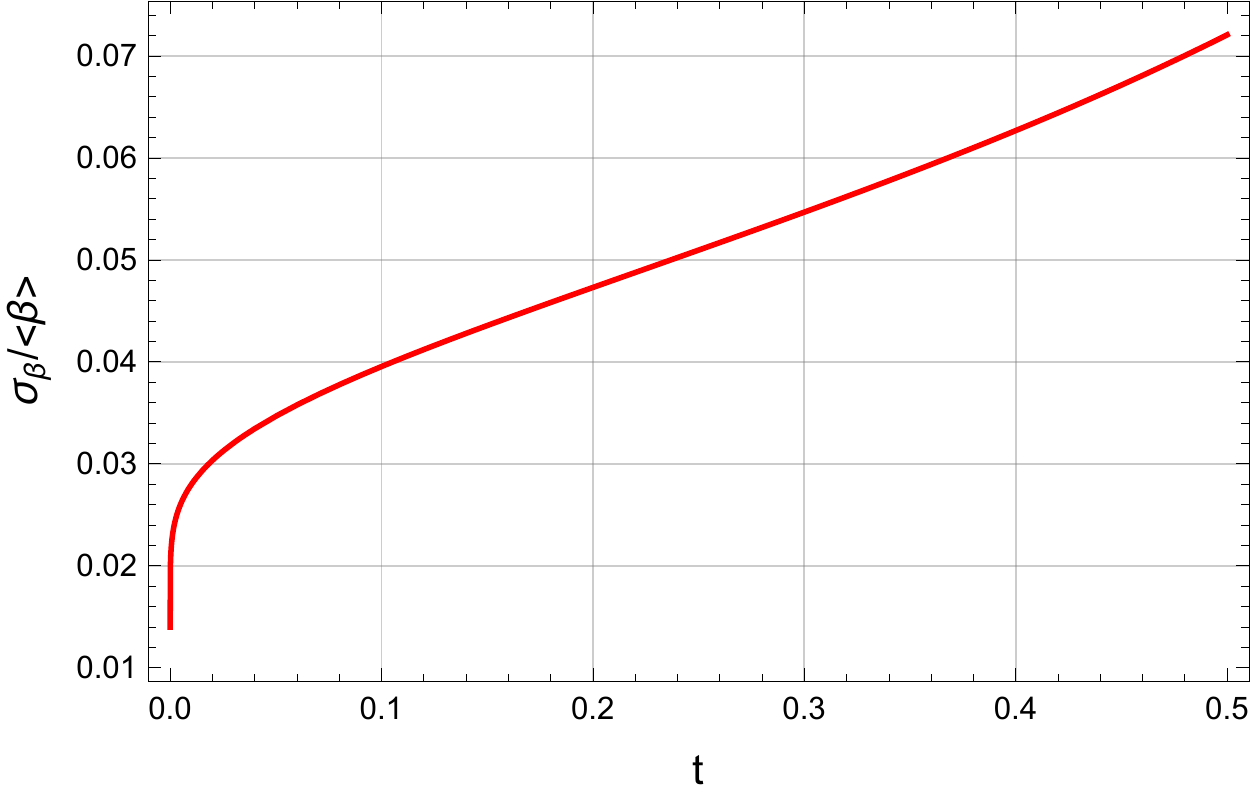}
\caption{\footnotesize The evolution of the ratio $\frac{\sigma_{\beta}}{<\widehat{\beta}_{\pm}>}$ as a function of time $t$. The ratio becomes zero in the limit $t\rightarrow 0$, so the Universe approach the singularity ``classically''}
\label{fig3}
\end{figure}
As we can see in the FIG. \ref{fig3}, the condition $\frac{\sigma_{\beta}}{<\widehat{\beta}_{\pm}>} \ll 1$ remains always valid during the approach to the singularity, and furthermore the ratio (\ref{ratio stan}) goes to zero  in the limit $t\rightarrow0$. Such occurrence tells us that the divergent behavior of the expectation values $<\widehat{\beta}_{\pm}>$  are always more and more ``classically ensured'' as we approach the singularity.

An additional confirm on the dynamics of the anisotropies can be provided by studying the behavior of the maximum of the wave packet built from the wave function (\ref{funz onda qua a}), in this way:
\begin{equation}
\label{pacc stan}
\Psi_{k_{\pm}^{*}}(t,\beta_{\pm}) = \int\int dk_{\pm}e^{-\frac{(k_{+}-k_{+}^{*})^{2}}{2\sigma_{+}^{2}}}e^{-\frac{(k_{-}-k_{-}^{*})^{2}}{2\sigma_{-}^{2}}}\varphi(t,\beta_{\pm}),
\end{equation}
where we choose Gaussian weights to peak the wave packets.
The evolution of the wave packets has been studied through a numerical integration and as we can see in the FIG.\ref{fig1}, the position of the maximum of the wave packet $|\Psi_{k_{\pm}^{*}}(t,\beta_{\pm})|$ (the collection of the black points in the Figure) as a function of $t$ overlaps exactly the trajectory of the anisotropies obtained by the Ehrenfest theorem in the Eq.(\ref{sol beta stan}).

\section{Polymer approach to the Bianchi I model in presence of a stiff matter contribution}
\label{sec:4}
As we have shown in the Section \ref{sec:2} for the one-dimensional particle case, if we consider the position variable $q$ as a discrete variable for the conjugated momenta $p$ it is not possible to associate a differential quantum operator. Thus, characterizing the wave function in the momentum polarization, through the polymer procedure we can identify an approximate version of the operator $\widehat{p}$ which acts multiplicatively on the states of the system.

For what concerns the Bianchi I model analyzed in the Section \ref{sec:3}, we make now the physical choice to assign a discrete character to the isotropic variable $a$ without modification in the anisotropy variables. Such a mixed choice for the quantization can be made typically in two cases: when the configuration variables are totally independent or when a configuration variable depends from the other one but weakly. The latter is exactly our case, where the variables $\beta_{\pm}$ depend parametrically only on the isotropic variable and so the effects of the variable $a$ on the anisotropies of the Universe are negligible. 

This means, following Eq.(\ref{appP2}), to consider the substitution
\begin{equation}
\label{sost poly}
p_{a}^{2} \rightarrow \frac{2 \hbar^{2}}{\lambda^{2}}\left[1-\cos\left( \frac{\hbar p_{a}}{\lambda} \right) \right].
\end{equation}
Since the results of the quantization procedure does not depend on the particular choice of the polarization, as it is clear from the Section \ref{sec:3} in our model, we choose to describe the wave function of the Universe $\psi=\psi(p_{a},\beta_{\pm})$ in the momentum base for the isotropic part and in the position base for the anisotropies.
This implies that the action of the operators are the same in Eq.(\ref{azione operatori p}) but taking into account the multiplicative action of the quantum operator associated to approximate version provided in the Eq.(\ref{sost poly}).
It brings to the modification of the superHamiltonian constraint (\ref{hamiltoniana}) and the WDW equation (\ref{WDW 3}) in this way:
\begin{widetext}
\begin{equation}
\label{hamiltoniana poly}
\mathcal{H}_{p} = \frac{l_{p}^{2}}{24 \pi \hbar}\left\{ -\frac{2 \hbar^{2}}{\lambda^{2}a}\left[1-\cos\left( \frac{\lambda p_{a}}{\hbar} \right) \right] + \frac{p_{+}^{2} + p_{-}^{2}}{a^{3}}\right\} + \frac{8 \pi^{2} \mu^{2}}{\hbar a^{3}} = 0
\end{equation}
\begin{equation}
\label{WDW poly}
\left\{ \hbar^{2}\frac{\partial}{\partial p_{a}}\left(\frac{2 \hbar^{2}}{\lambda^{2}}\left[1-\cos\left( \frac{\lambda p_{a}}{\hbar} \right) \right]\frac{\partial}{\partial p_{a}}\right) - \hbar^{2}\left(\frac{\partial^{2}}{\partial \beta_{+}^{2}} + \frac{\partial^{2}}{\partial \beta_{-}^{2}}\right) +\frac{3 (4 \pi)^{3}\mu^{2}}{l_{p}^{2}} \right\} \psi(p_{a},\beta_{\pm})=0
\end{equation} 
\end{widetext}

From the Eq.(\ref{WDW poly}) it is possible to obtain a modified polymer current as
\begin{widetext}
\begin{displaymath}
\label{corr poly p}
J^{\mu} =
\left[ \begin{array}{ccc}
J^{p_{a}} \\
J^{+} \\
J_{-}
\end{array} \right] = -\frac{i}{2}\hbar^{2} \left[ \begin{array}{ccc}
 \frac{2 \hbar^{2}}{\lambda^{2}}\left[1-\cos\left( \frac{\lambda p_{a}}{\hbar} \right) \right] ( \psi^{*} \partial_{p_{a}}\psi - \psi \partial_{p_{a}}\psi^{*}) \\
-( \psi^{*} \partial_{+}\psi - \psi \partial_{+}\psi^{*})  \\
-( \psi^{*} \partial_{-}\psi - \psi \partial_{-}\psi^{*})
\end{array} \right],
\end{displaymath}
\end{widetext}
with the associated conservation law $\nabla_{i}J^{i}=0$.
This case, the superspace metric is $h^{ij}=diag\left\{\hbar^{2}\frac{2 \hbar^{2}}{\lambda^{2}}\left[1-\cos\left( \frac{\hbar p_{a}}{\lambda}\right) \right],-\hbar^{2},-\hbar^{2} \right\}$ and being $\Gamma^{p_{a}}_{p_{a}p_{a}} = -\frac{\lambda}{\hbar}\frac{\sin\left(\frac{\lambda p_{a}}{\hbar}\right)}{1 - \cos \left( \frac{\lambda p_{a}}{\hbar} \right)}$  the only non-vanishing Christoffel symbol, we can evaluate the explicit form of the conserved current as
\begin{equation}
\label{conservazione esplicita poly}
\nabla_{i}J^{i}= \left( \partial_{p_{a}} - -\frac{\lambda}{\hbar}\frac{\sin\left(\frac{\lambda p_{a}}{\hbar}\right)}{1 - \cos \left( \frac{\lambda p_{a}}{\hbar} \right)}\right)J^{p_{a}} + \partial_{+}J^{+} + \partial_{-}J^{-}=0.
\end{equation}

\subsection{Semiclassical Limit}
What we want to realize is the  implementation of the Vilenkin approach for the polymer version of the WDW equation. In other words, we consider the wave function of the Universe (\ref{funz p}) in the Eq.(\ref{WDW poly}). At the lowest order of the expansion in $\hbar$, namely the semiclassical level, the Hamilton-Jacobi equation obtained can be written as
\begin{equation}
\label{impulso poly}
p_{a}^{2} = \frac{\hbar^{2}}{\lambda^{2}}\arccos\left(  1 - \frac{3 (4\pi)^{3}\mu^{2}\lambda^{2}}{2 \hbar^{2}l_{p}^{2}a^{2}}\right)^{2},
\end{equation}
where we have identified again $\dot{S} = a$.
From the superHamiltonian (\ref{hamiltoniana poly}) we can write the Hamiltonian equation for the isotropic variable:
\begin{equation}
\label{Hamilton poly}
\frac{d a}{d t} = \frac{\partial \mathcal{H}_{p}}{\partial p_{a}} = -\frac{l_{p}^{2}}{12 \pi \lambda a }\sin\left( \frac{\lambda p_{a}}{\hbar} \right)
\end{equation}
Then, we introduce the the Eq.(\ref{impulso poly}) in the Eq.(\ref{Hamilton poly}) using the trigonometric relation $\sin (\arccos (x))=\sqrt{1-x^{2}}$ so to obtain
\begin{equation}
\label{diff poly}
\frac{d a}{d t} = \sqrt{\frac{4 \pi}{3}}\frac{l_{p} \mu}{\hbar a^{2}}\sqrt{1 - \frac{48 \pi^{3} \mu ^{2} \lambda^{2}}{\hbar^{2}l_{p}^{2} a^{2}}}.
\end{equation}
Looking at the latter equation it is immediate to see the existence of a particular value
\begin{equation}
\label{minimo}
a_{min}= \sqrt{\frac{48 \pi^{3} \mu ^{2} \lambda^{2}}{\hbar^{2}l_{p}^{2} }}
\end{equation}
for which $\frac{d a}{d t}$ changes the sign, or in other words a stationary point for the function $a(t)$.
The branch of the solution that we are interested to compare with the standard behavior (\ref{diff1}) can be obtained through an analytic integration of the differential equation (\ref{diff poly}) and its form is:
\begin{widetext}
\begin{multline}
\label{a poly}
a(t)=\frac{1}{\hbar l_{p}}\sqrt{ 6 \pi \mu ^2 \left[h^2 l^4 t \left(h^2 l^4 t +
   \sqrt{h^4 l^8 t^2+36864 \pi ^8 \lambda ^6 \mu ^4}\right)+18432 \pi ^8 \lambda ^6
   \mu ^4\right]^\frac{1}{3}} + \\ \overline{+48 \pi ^3 \lambda ^2 \mu ^2 \left[\left(\frac{72 \pi ^{8}\lambda ^6 \mu ^{4}}{h^2 l^4 t \left(h^2 l^4 t +
   \sqrt{h^4 l^8 t^2+36864 \pi ^8 \lambda ^6 \mu ^4}\right)+18432 \pi ^8 \lambda ^6
   \mu ^4}\right)^\frac{1}{3}-1\right]},
\end{multline}
\end{widetext}
where we have chosen the integration in such a way that the stationary point $a_{min}$ is reached in correspondence to $t=0$.
\begin{figure}[h!]
\label{ciccio}
\includegraphics[scale=.67]{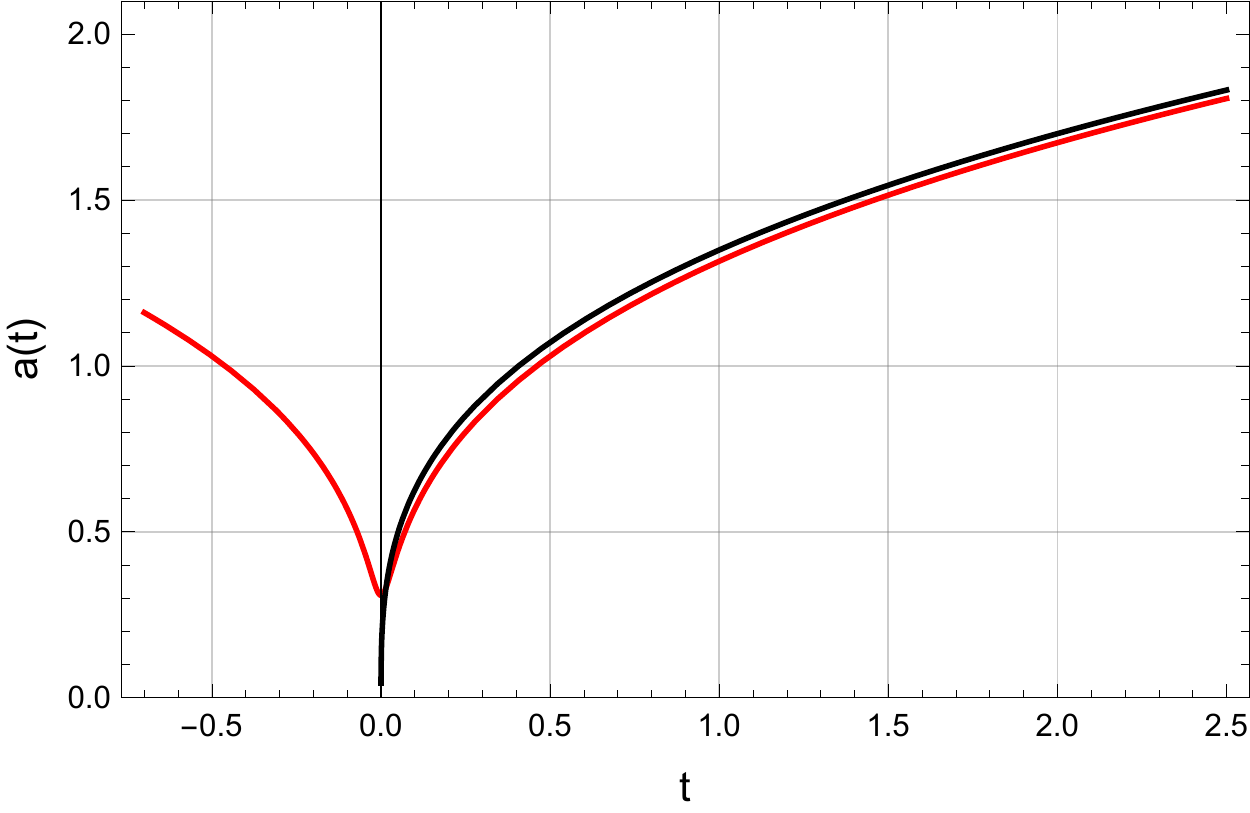}
\caption{The black line represents the standard behavior $a(t)$ as evaluated in the Eq.(\ref{diff1}) and the red line represents the polymer behavior of the isotropic variable (\ref{a poly}). The solution is sketched for the parameters: $\hbar = \l_{p} = 1, \lambda = 0.01, \mu = 0.4$. The standard solution reaches the singularity in $t=0$ while the polymer solution arrives at the minimum value $a_{min}$ and then grows up for $t<0$ after the bounce.}
\label{fig2}
\end{figure}

As it is possible to see in the FIG. \ref{fig2}, the stationary point is associated to a minimum value of the variable $a(t)$. Differently from the standard case, in the presence of a polymer structure, the isotropic variable does not reach $a=0$ in correspondence of $t=0$ and furthermore a collapsing phase towards the singularity is followed by a contracting phase. Let us emphasize that the obtained solution reproduce the standard results, formally it reduces to the solution (\ref{diff1}), in the regimes where we expect that a polymer modification does not change the dynamics. In fact, in the limit $\lambda \rightarrow 0 $ the expression (\ref{a poly}) assumes the form (\ref{diff1}) and for late times (namely $t->+\infty$) the two solutions tend to be indistinguishable.

The presence of the small parameter $\lambda\neq 0$ within the theory associated to the lattice on the isotropic variable has led to a cosmological model in which the initial singularity of the big bang has been avoided and it has been replaced with a bounce.

\subsection{Quantum subsytem}

In this subsection we analyse the first order in $\hbar$ obtained considering the wave function (\ref{funz onda qua p}) in the WDW equation (\ref{WDW poly}). This brings to the equation:
\begin{widetext}
\begin{equation}
\label{primo ordine p poly}
i \frac{2 \hbar^{2}}{\lambda^{2}}\left[1-\cos\left( \frac{\lambda p_{a}}{\hbar} \right) \right] \frac{1}{A}(A^{2}\dot{S})^{\dot{}} + 2 i \frac{2 \hbar^{2}}{\lambda^{2}}\left[1-\cos\left( \frac{\lambda p_{a}}{\hbar} \right) \right] A \dot{S} \dot{\varphi} - A \left(\partial_{+}^{2} + \partial_{-}^{2}\right)\varphi = 0.
\end{equation} 
\end{widetext}

Making use of the adiabatic approximation (\ref{app adia p}), the above expression provides, as in the standard case, a pair of equations. The first one is
\begin{equation}
\label{cons corrente poly classica}
i \frac{2 \hbar^{2}}{\lambda^{2}}\left[1-\cos\left( \frac{\lambda p_{a}}{\hbar} \right) \right] \frac{1}{A}(A^{2}\dot{S})^{\dot{}} = 0
\end{equation}
and it concerns the conservation of the classical current $\nabla_{p_{a}} J^{p_{a}} = 0$ which explicit form in the polymer case and with the wave function (\ref{funz p}) is: 
\begin{equation}
\label{corr poly classica}
J^{p_{a}} =-\hbar\frac{2 \hbar^{2}}{\lambda^{2}}\left[1-\cos\left( \frac{\lambda p_{a}}{\hbar} \right) \right]A^{2}\dot{S}.
\end{equation}
The second equation gives the description of the quantum part of the wave function $\varphi$:
\begin{equation}
\label{quantistica poly}
 2 i \frac{2 \hbar^{2}}{\lambda^{2}}\left[1-\cos\left( \frac{\lambda p_{a}}{\hbar} \right) \right]\dot{S} \dot{\varphi} =\left(\partial_{+}^{2} + \partial_{-}^{2}\right)\varphi 
\end{equation}
It is possible to rewrite also in the polymer scheme a pure Schrodinger equation. First of all, from the H-J equation (\ref{impulso poly}) we achieve the expression for the trigonometric term:
\begin{equation}
\label{trig}
\frac{2 \hbar^{2}}{\lambda^{2}}\left[1-\cos\left( \frac{\lambda p_{a}}{\hbar} \right) \right] = \frac{3 (4\pi)^{3}\mu^{2}}{l_{p}^{2}a^{2}}.
\end{equation}
Moreover, we can use the relations $\dot{S} = a$ and $ \dot{\varphi} = \frac{\partial \varphi}{\partial t}\frac{\partial t}{\partial p_{a}} = \frac{\partial \varphi}{\partial t}\frac{1}{\dot{p_{a}}}$.
The last step is to evaluate $\dot{p_{a}}$ through a  differentiation of the relation (\ref{impulso poly}) and the Eq.(\ref{diff poly}) in order to write
\begin{equation}
\label{p punto poly}
\dot{p_{a}} = \frac{16 \pi^{2} \mu^{2}}{\hbar a^{4}}
\end{equation}
Realizing all the previous substitutions the Eq.(\ref{quantistica poly}) reduces to 
\begin{equation}
\label{schro poly}
i \left(\frac{24 \pi \hbar}{l_{p}^{2}}\right)a^{3}\frac{\partial \varphi (t, \beta_{\pm})}{\partial t} =  \left(\partial_{+}^{2} + \partial_{-}^{2}\right)\varphi(t, \beta_{\pm}).
\end{equation}
It is important to note that the functional form of the equation that describes the quantum subsystem is exactly the same of the standard case in the Eq.(\ref{schro 2}). The real difference is in the time-dependence of the isotropic factor $a$ that, in the polymer case, assumes the form in the Eq.(\ref{a poly}). For this reason, we proceed defining the same change in the time variable $\frac{\partial }{\partial \tau} = \frac{24 \pi}{l_{p}^{2}}a^{3}\frac{\partial }{\partial t}$ in order to arrive at the same Schrodinger equation (\ref{schro 1}) with solution (\ref{onda piana}).
With respect to the standard case the distinction is in the integration of the time-like variable $\tau$
\begin{equation}
\label{int tau poly}
\tau(t)=\int \frac{l_{p}^{2}}{24\pi} \frac{dt}{a(t)^{3}}.
\end{equation} 
In fact, it brings to a non-solvable integral  in the $t$ variable due to the special form $a(t)$ in the polymer case.
We can elude this changing the integration variable this way:
\begin{equation}
\label{int tau solvable}
d \tau = \frac{l_{p}^{2}}{24 \pi}\frac{dt}{a^{3}} = \frac{l_{p}^{2}}{24 \pi}\frac{1}{\dot{a}}\frac{da}{a^{3}}
\end{equation}
The expression $\tau(a)$ can be determined integrating the latter equation and considering the relation (\ref{diff poly}) in order to obtain:
\begin{equation}
\tau(a) = \frac{\sqrt{3}l_{p}\hbar}{48 \pi^{3/2}\mu}\int \frac{da}{\sqrt{a^{2}- \frac{48 \pi^{3} \mu ^{2} \lambda^{2}}{\hbar^{2}l_{p}^{2} }}},
\end{equation} 
which admits the analytic solution
\begin{equation}
\label{soluzione per tau}
\tau(a(t)) = \frac{\sqrt{3}l_{p}\hbar}{48 \pi^{3/2}\mu}\log\left[ \frac{\left( a(t)+\sqrt{a(t)^{2} - \frac{48 \pi^{3} \mu ^{2} \lambda^{2}}{\hbar^{2}l_{p}^{2} } } \right)}{\left( a(t^{*})+\sqrt{a(t^{*})^{2} - \frac{48 \pi^{3} \mu ^{2} \lambda^{2}}{\hbar^{2}l_{p}^{2} } } \right) }\right].
\end{equation}
Certainly, if we implement the limit $\lambda\rightarrow0$ and we substitute the standard time dependence of the isotropic variable (\ref{diff1}) we turn back to the expression (\ref{tau a}).
This allow to write down the analytic version of the quantum part of the wave function $\varphi$:
\begin{equation}
\label{funz onda qua poly}
\varphi(t,\beta_{\pm}) = Ce^{ i(k_{+}^{2} + k_{-}^{2})\tau\left(a(t)\right)}e^{\frac{ik_{+}\beta_{+}}{\hbar}}e^{\frac{ik_{-}\beta_{-}}{\hbar}}.
\end{equation}

To find the probability distribution for the polymer Bianchi I model in presence of a stiff matter contribution, we reply the steps of the Section \ref{3.1}. For what concern the classical part of the probability distribution, the situation is the same of the standard case. Indeed, the presence of just one classical configuration variable, $p_{a}$, denotes the impossibility to recast the conservation of the classical current (\ref{cons corrente poly classica}) in a continuity equation that regards the classical probability distribution. However, also in the polymer case, for the quantum sub-system a continuity equation can be extracted. Referring to the quantum part of the wave function (\ref{funz onda qua poly}], the probability distribution for the quantum variables is defined as $\rho_{\varphi}=|\varphi|^{2}$. Considering the Vilenkin wave function (\ref{funz onda qua poly}), the leading terms of the components of the current (\ref{WDW poly}) become
\begin{equation}
\label{clas a poly}
J^{a} =\hbar\frac{2 \hbar^{2}}{\lambda^{2}}\left[1-\cos\left( \frac{\lambda p_{a}}{\hbar} \right) \right]A^{2}\dot{S}\rho_{\varphi},
\end{equation}
\begin{equation}
\label{clas beta poly}
J^{\pm} = -\frac{\hbar^{2}A^{2}}{2}( \varphi^{*}\partial_{\pm}\varphi - \varphi\partial_{\pm}\varphi^{*} ) \equiv \frac{A^{2}}{2}J^{\pm}_{\varphi},
\end{equation}
and the conservation law $\nabla_{\mu}J^{\mu}=0$ can be recast as
\begin{equation}
\label{eq cont poly}
2\hbar \frac{2 \hbar^{2}}{\lambda^{2}}\left[1-\cos\left( \frac{\lambda p_{a}}{\hbar} \right) \right] \dot{S}\frac{d\varphi}{d p_{a}} + \partial_{i}J^{i}_{\varphi}=0,
\end{equation}
where the index $i=\{+,-\}$. Via the relation $\frac{\partial \varphi}{\partial p_{a}} = \frac{\partial \varphi}{\partial t}\frac{\partial t}{\partial p_{a}} =\frac{\partial \varphi}{\partial t}\frac{1}{\dot{p}_{a}}$, the explicit dependence on the variable $t$ can be inserted. Furthermore, considering the Hamilton-Jacobi equation (\ref{HJ 3}), the identification $\dot{S}=a$ and the identity (\ref{p punto poly}), the Eq.(\ref{eq cont poly}) reduces to
\begin{equation}
\label{continuita a}
\frac{d \rho_{\varphi}}{dt} = -\frac{l_{p}^{2}}{24 \hbar^{2}\pi a^{3}(t)}\partial_{i}J^{i}_{\varphi}.
\end{equation}
The continuity equation obtained in the polymer case is formally equivalent with respect to the standard case (\ref{continuita a}), whose differences are due to the time dependence of the isotropic variable $a$ (corresponding to the Eq.(\ref{a poly}) in the polymer case) and
to the definition of the quantum probability distribution given by the quantum wave function (\ref{funz onda qua poly}) with respect to the standard case (\ref{funz onda qua p}). That said, performing again an integration over a $d\beta_{+}d\beta_{-}$ volume for both sides of the continuity equation, a normalizable quantum probability distribution is obtained also in the polymer case:
\begin{equation}
\label{normalizzazione poly}
\int \int d\beta_{+}d\beta_{-}\rho_{\varphi} = 1
\end{equation}

In conclusion of this section we assert that it was not possible to build an entire conserved probability distribution as in the Eq.(\ref{normalizzazione intera}) but, as in the standard case, we were able to write a quantum normalizable probability distribution (\ref{normalizzazione poly}) together with a classical conserved quantity (\ref{cons corrente poly classica}).

\subsection{Consequences on the anisotropies behavior}

We investigate the behavior of the anisotropies trough the evaluation of the quantum expectation values. As in Section \ref{sec:3}, the application of the Ehrenfest Theorem to $\widehat{\beta_{\pm}}$ brings to the same functional form also in the Polymer case:
\begin{equation}
\label{ehren poly}
\frac{d<\widehat{\beta}_{\pm}>}{dt} = \frac{1}{i \hbar}<[\beta_{\pm},\mathcal{H}_{p}]>  = \frac{l_{p}^{2}p_{\pm}}{12 \pi \hbar a^{3}},
\end{equation}
where in this case the variable $a$ behaves as in the Eq.(\ref{a poly}).
The previous differential equation can be solved by changing variable in the following way:
\begin{equation}
\label{ehren poly 2}
\frac{d<\widehat{\beta}_{\pm}>}{da} = \frac{l_{p}^{2}p_{\pm}}{12 \pi \hbar a^{3}\dot{a}} = \frac{l_{p}p_{\pm}}{\sqrt{192 \pi^{3} } \mu}\frac{1}{\sqrt{a^{2} - a^{2}_{min}}},
\end{equation}
and the solution obtained is:
\begin{equation}
\label{sol beta stan poly}
<\widehat{\beta}_{\pm}>_{a(t)} = \frac{l_{p} p_{\pm}}{\sqrt{192 \pi^{3}} \mu}\ln\left[\frac{a(t)^{2} + \sqrt{a(t)^{2} - a^{2}_{min}}}{a(t^{*})^{2} + \sqrt{a(t^{*})^{2} - a^{2}_{min}}}  \right].
\end{equation}
The integration constant has been chosen to reproduce, in the limit $\lambda \rightarrow 0$, the solutions of the standard case Eq.(\ref{sol beta stan}).	
\begin{figure}[h!]
\includegraphics[scale=.67]{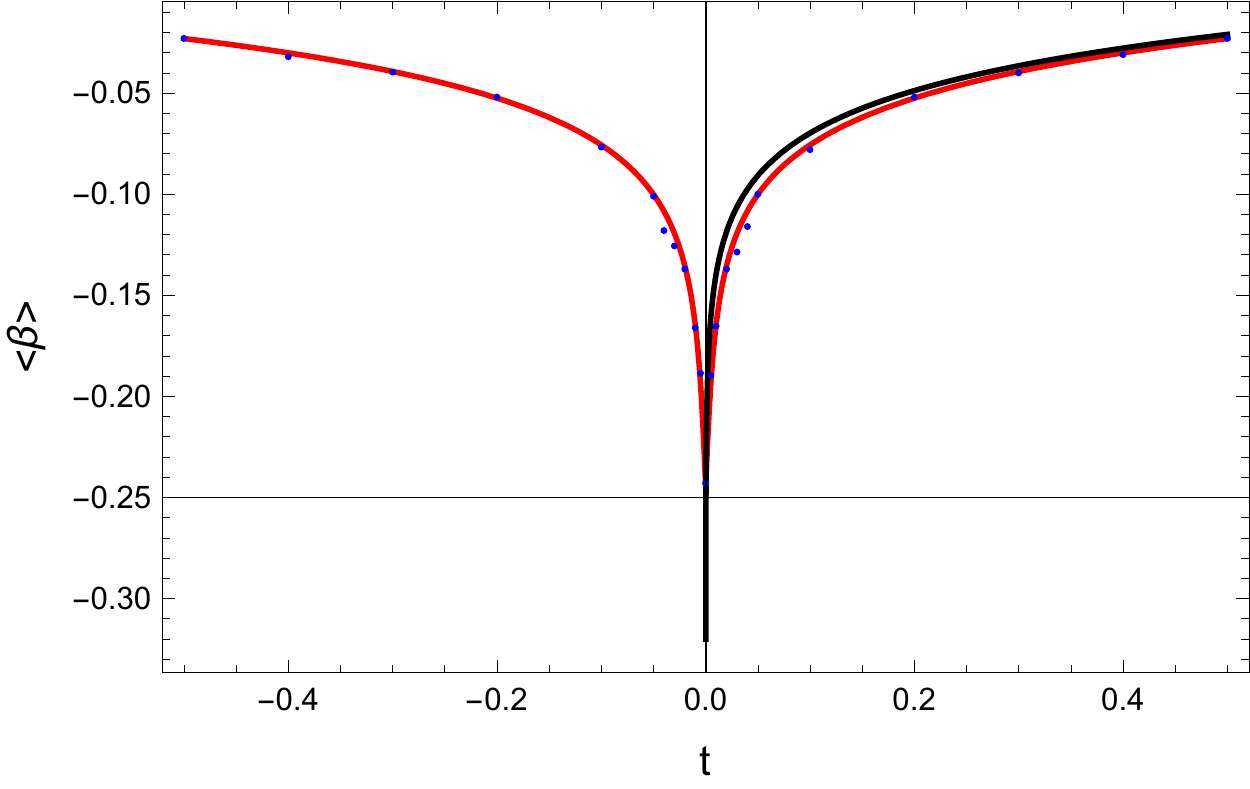}
\caption{\footnotesize The black trajectory represents the standard divergent behavior of the anisotropies, as obtained through the Ehrenfest theorem in the Eq.(\ref{sol beta stan}). The red trajectories shows the finite value that the anisotropies assume in the turning point. Then, the blue points stands for the position of the maximum of the wave packets (\ref{pacc poly}). The equivalence in the consideration of the Ehrenfest treatment and the wave packet dynamics is ensured in the total overlap between red trajectory and blue points.}
\label{fig4}
\end{figure}

In FIG. \ref{fig4} is shown the behavior of the quantum expectation value $<\widehat{\beta_{\pm}}>$ in the polymer case (red trajectory) and in the standard case (black trajectory). First of all, for the two trajectories there is an overlapping in the limit $t\rightarrow\infty$, so for late times there are no polymer modification. Furthermore, the divergent character close to the initial singularity shown by the solution in the standard case disappears leaving the place to a minimum point (or a maximum depending on the relative sign between the parameters $p$ and $\mu$) corresponding to
\begin{equation}
\label{beta star}
<\widehat{\beta}_{\pm}>^{*} = \frac{l_{p} p_{\pm}}{\sqrt{192 \pi^{3}} \mu}\ln\left[\frac{a^{2}_{min}}{a(t^{*})^{2} + \sqrt{a(t^{*})^{2} - a^{2}_{min}}}  \right].
\end{equation}
The anisotropies cross over the singularity remaining finite assuming, in correspondence of $t=0$, the value $<\widehat{\beta}_{\pm}>^{*}$ that depends on the choice of the parameters and the initial conditions $t^{*}$. 
We stress how such a meaningful achievement relies on the assumptions at the ground of our dynamical study, \textit{i.e.} a WKB semiclassical limit, the presence of stiff matter and overall the implementation of a polymer discretization paradigm. The latter two requirements both contribute  to obtain a cut-off dynamics on the Universe asymptotic evolution, which in turn ensures the anisotropies finiteness. It is worth empathizes that relaxing the semiclassical approximation to the Universe wave function can permit the anisotropies blow up, see for instance the recent study in \cite{barrow}.

As in the standard case, the evaluation of the standard deviation is a good tool to appreciate if the evolution (\ref{sol beta stan poly}) it is not so different from the classical trajectory.

The Ehrenfest Theorem for the operator $<\widehat{\beta}^{2}>$ in the polymer case leads to:
\begin{multline}
\label{ehren poly posi quad}
\frac{d<\widehat{\beta}_{\pm}^{2}>}{dt} = \frac{1}{i \hbar}<[\beta_{\pm}^{2},\mathcal{H}_{p}]> =\frac{l_{p}^{2}<[\beta_{\pm}^{2},p_{\pm}^{2}]>}{24 i \pi \hbar^{2} a^{3}} = \\ = \frac{l_{p}^{2}}{12 \pi \hbar a^{3}}<\beta_{\pm}p_{\pm} + p_{\pm}\beta_{\pm}>,
\end{multline}
where, differently from the Eq.(\ref{ehren stan posi quad}), the isotropic variable $a$ concerns the Eq.(\ref{a poly}).
A new application of the Ehrenfest theorem on the quantity $<\beta_{\pm}p_{\pm} + p_{\pm}\beta_{\pm}>$ allow to obtain a differential equation for this term:
\begin{multline}
\frac{d <\beta_{\pm}p_{\pm} + p_{\pm}\beta_{\pm}>}{dt} = \\ = \frac{l_{p}^{2}<[\beta_{\pm}p_{\pm} + p_{\pm}\beta_{\pm},p_{\pm}^{2}]>}{24 i \pi \hbar^{2} a^{3}} =\frac{l_{p}^{2}p_{\pm}^{2}}{6 \pi \hbar a^{3}},
\end{multline} 
whose solution can be achieve with a change of variable such that:
\begin{equation}
\frac{d <\beta_{\pm}p_{\pm} + p_{\pm}\beta_{\pm}>}{da} = \frac{l_{p}^{2}p_{\pm}^{2}}{6 \pi \hbar a^{3}\dot{a}} = \frac{l_{p}p_{\pm}^{2}}{4\pi \sqrt{3 \pi}\mu}\frac{1}{\sqrt{a^{2} - a^{2}_{min}}},
\end{equation}
and it corresponds to
\begin{multline}
\label{sol intermedia poly}
<\beta_{\pm}p_{\pm} + p_{\pm}\beta_{\pm}> = \\ =\frac{l_{p}p_{\pm}^{2}}{4\pi \sqrt{3 \pi}\mu}\left(\ln\left[a(t)^{2} + \sqrt{a(t)^{2} - a^{2}_{min}}  \right] + C\right),
\end{multline}
where $C$ is an integration constant.
\begin{figure}[h!]
\includegraphics[scale=.67]{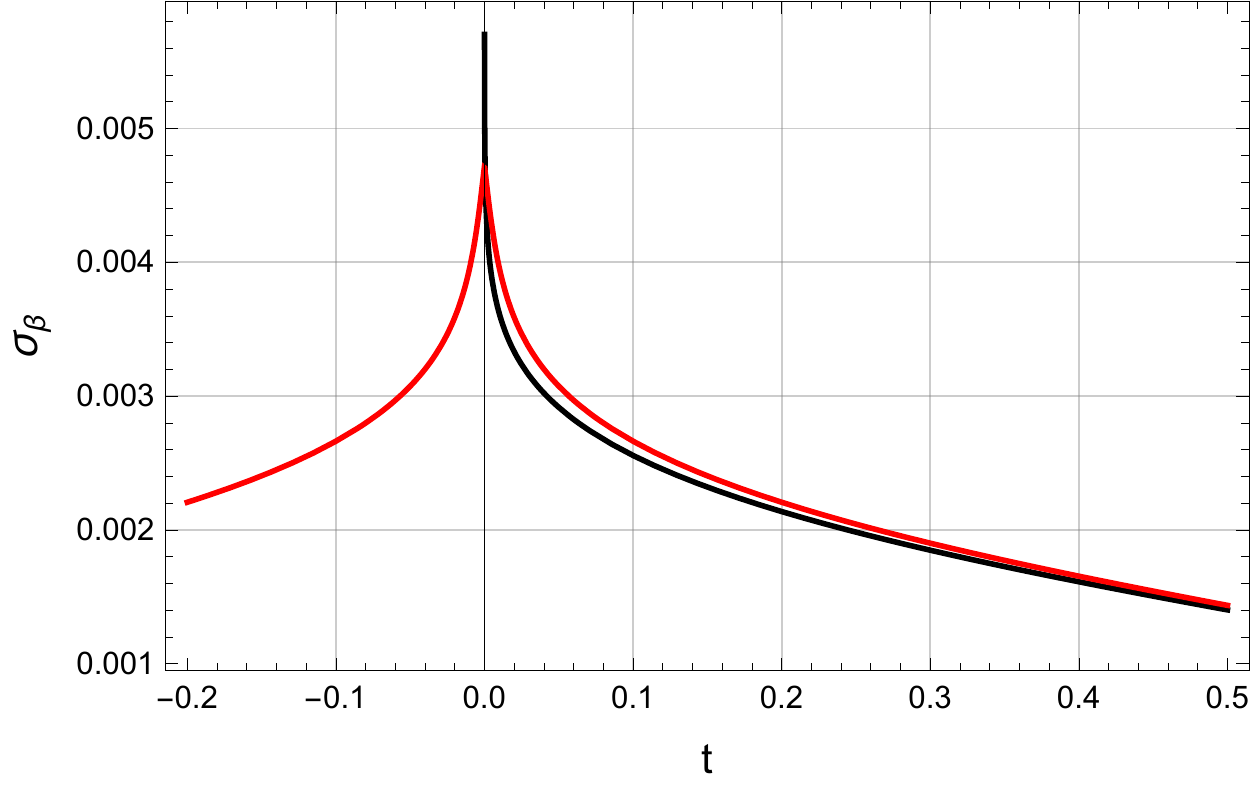}
\caption{\footnotesize A comparison between the standard deviation in the canonical case (\ref{dev stan})(black) and in the polymer case (\ref{dev stan poly})(red). A regularization for the standard deviation in correspondence of the turning point emerges in the polymer scheme.}
\label{fig5}
\end{figure}
\begin{figure}[h!]
\includegraphics[scale=.67]{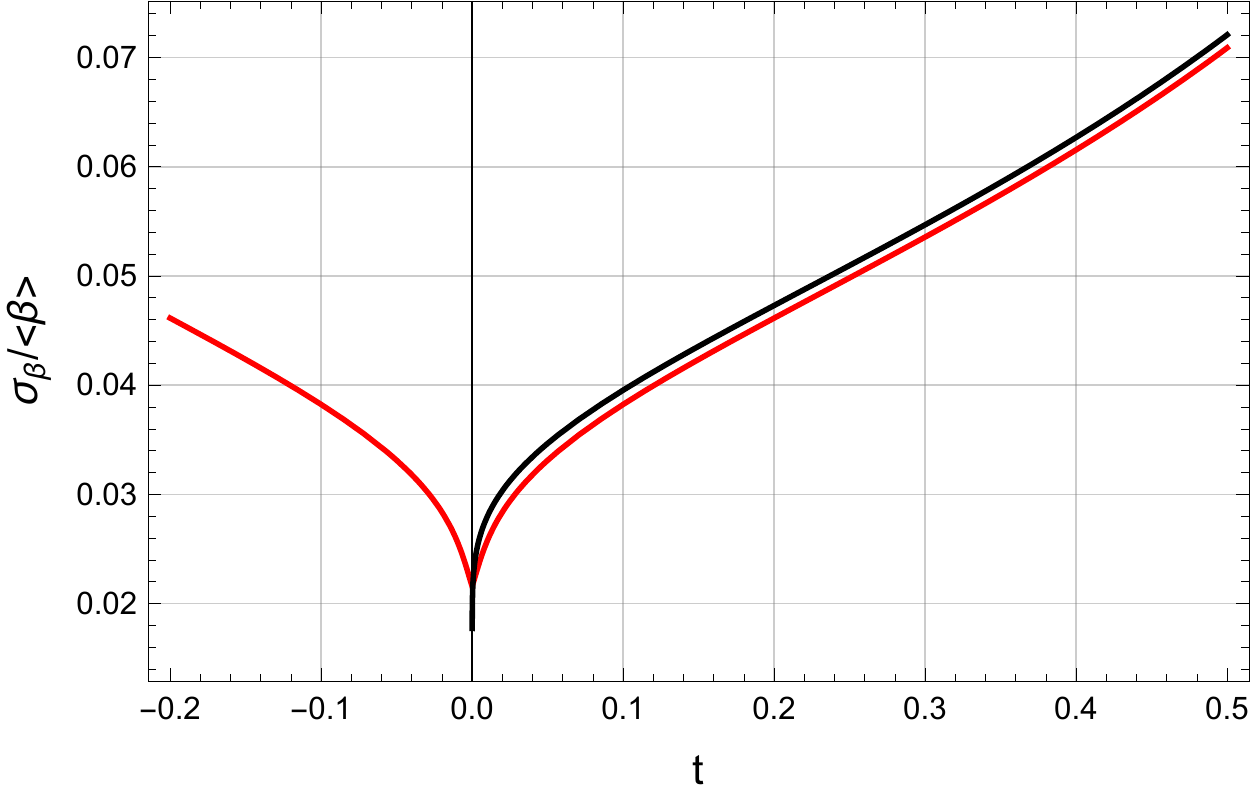}
\caption{\footnotesize A comparison between the ratio $\frac{\sigma_{\beta}}{<\widehat{\beta}_{\pm}>}$ in the canonical case (\ref{ratio stan})(black) and in the polymer case (red). In the polymer scheme this ratio remains finite in correspondence of the turning point.}
\label{fig6}
\end{figure}

Using the same change of variable of the Eq.(\ref{ehren poly posi quad}) we can arrive to a treatable form of the differential equation for the  expectation value $<\widehat{\beta}_{\pm}^{2}>$ with the correspondent analytic solution:
\begin{widetext}
\begin{equation}
\label{diff beta quadro poly}
\frac{d<\widehat{\beta}_{\pm}^{2}>}{da}= \frac{l_{p}^{2}p_{\pm}^{2}}{96\pi^{3}\mu^{2}}\frac{1}{\sqrt{a^{2} - a^{2}_{min}}}\left(\ln\left[a(t)^{2} + \sqrt{a(t)^{2} - a^{2}_{min}}  \right] + C \right).
\end{equation}
\begin{multline}
\label{sol beta quadro poly}
<\widehat{\beta}_{\pm}^{2}>_{a(t)}=\frac{l_{p}^{2}p_{\pm}^{2}}{192\pi^{3} \mu^{2}}\left(2C\ln\left[\frac{a(t)^{2} + \sqrt{a(t)^{2} - a^{2}_{min}}}{a(t^{*})^{2} + \sqrt{a(t^{*})^{2} - a^{2}_{min}}}  \right] \right. + \\ \left. + \ln^{2}\left[a(t)^{2} + \sqrt{a(t)^{2} - a^{2}_{min}}  \right] -  \ln^{2}\left[a(t^{*})^{2} + \sqrt{a(t^{*})^{2} - a^{2}_{min}}  \right]\right)
\end{multline}
\end{widetext}

Then, the standard deviation for the operator $<\widehat{\beta}_{\pm}>$ in the polymer case can be written following the usual definition in order to obtain:
\begin{widetext}
\begin{multline}
\label{dev stan poly}
\sigma_{\beta} = \sqrt{<\widehat{\beta}_{\pm}^{2}> -<\widehat{\beta}_{\pm}>^{2} }  = \frac{l_{p} p_{\pm}}{\sqrt{192 \pi^{3}} \mu}\sqrt{2\left(-\ln^{2}\left[a(t^{*})^{2} + \sqrt{a(t^{*})^{2} - a^{2}_{min}}  \right] +  C \ln\left[\frac{a(t)^{2} + \sqrt{a(t)^{2} - a^{2}_{min}}}{a(t^{*})^{2} + \sqrt{a(t^{*})^{2} - a^{2}_{min}}}  \right] \right.} + \\ \overline{\left. + \ln\left[a(t)^{2} + \sqrt{a(t)^{2} - a^{2}_{min}}  \right]\ln\left[a(t^{*})^{2} + \sqrt{a(t^{*})^{2} - a^{2}_{min}}  \right]\right)}
\end{multline}
\end{widetext}
Here again, the presence of the square root in the definition of the standard deviation requires that,in order to have a real number associated to this quantity, the constant of integration $C$ can assumes only particular values. 
The first difference respect to the standard case is glaring in the FIG.\ref{fig5}, where the black line represents the standard deviation evaluated in the Eq.(\ref{dev stan}) while the red line is a representation of the modified equation (\ref{dev stan poly}). In the presence of the polymer modification the standard deviation does not diverge in correspondence of the bounce but reaches a finite maximum value. Moreover, also the analysis of the ratio $\frac{\sigma_{\beta}}{<\widehat{\beta}_{\pm}>}$ confirms that the expectation values (\ref{ehren poly}) is a genuine quantity. In fact, as it is shown in the FIG. \ref{fig6}, the condition $\frac{\sigma_{\beta}}{<\widehat{\beta}_{\pm}>} \ll 1$ remains valid throughout the time evolution of the anisotropies, including the crossing of the bounce.

In the polymer case too it is possible to obtain an additional confirm on the dynamics of the anisotropies  by studying the behavior of the maximum of the wave packet built from the wave function (\ref{funz onda qua poly}), in this way:
\begin{equation}
\label{pacc poly}
\Psi_{k_{\pm}^{*}}(t,\beta_{\pm}) = \int\int dk_{\pm}e^{-\frac{(k_{+}-k_{+}^{*})^{2}}{2\sigma_{+}^{2}}}e^{-\frac{(k_{-}-k_{-}^{*})^{2}}{2\sigma_{-}^{2}}}\varphi(t,\beta_{\pm}),
\end{equation}
where we choose Gaussian weights to peak the wave packets.
A numerical integration has been realized to evaluate the evolution of the wave packets when the turning point is approached. As we can see in Fig. \ref{fig4}, the position of the maximum of the wave packet $|\Psi_{k_{\pm}^{*}}(t,\beta_{\pm})|$ as a function of $t$(correspondent to the collection of blue points) overlaps exactly the polymer trajectory founded by the Ehrenfest theorem in the Eq.(\ref{sol beta stan poly}).

\section{Implication on the Bianchi IX Model}
\label{sec:5}

The purpose of this Section is to implement the proprieties founded before to a more general model. For this reason, we take into account a Bianchi IX model filled with a stiff matter considering the same polymer prescription in the Eq.(\ref{sost poly}) for the isotropic variable. The
superHamiltonian constraint express through the configurational variables $ \{a, \beta_{+}, \beta_{-} \}$ takes the form
\begin{multline}
\label{hamiltoniana BIX}
\mathcal{H} = \frac{l_{p}^{2}}{24 \pi \hbar}\left\{ -\frac{2 \hbar^{2}}{\lambda^{2}a}\left[1-\cos\left( \frac{\lambda p_{a}}{\hbar} \right) \right] + \frac{p_{+}^{2} + p_{-}^{2}}{a^{3}} \right. + \\ + \left. \frac{12 \pi^{2}\hbar^{2}}{l_{p}^{4}}aV_{IX}(\beta_{\pm})\right\} + \frac{8 \pi^{2} \mu^{2}}{\hbar a^{3}} = 0,
\end{multline}
\begin{widetext}
\begin{figure*}[tb]
\centering
\includegraphics[scale=.45]{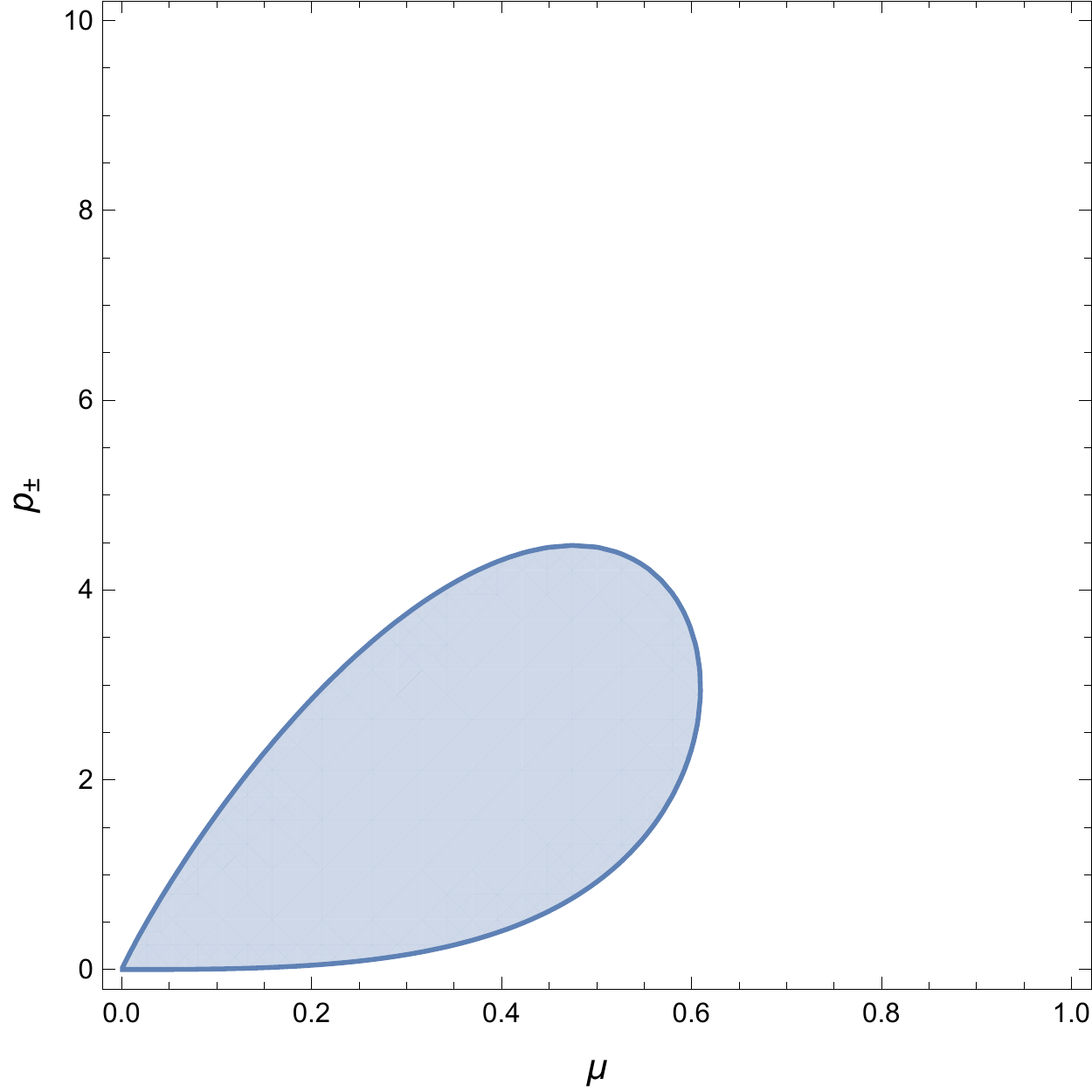}
\includegraphics[scale=.46]{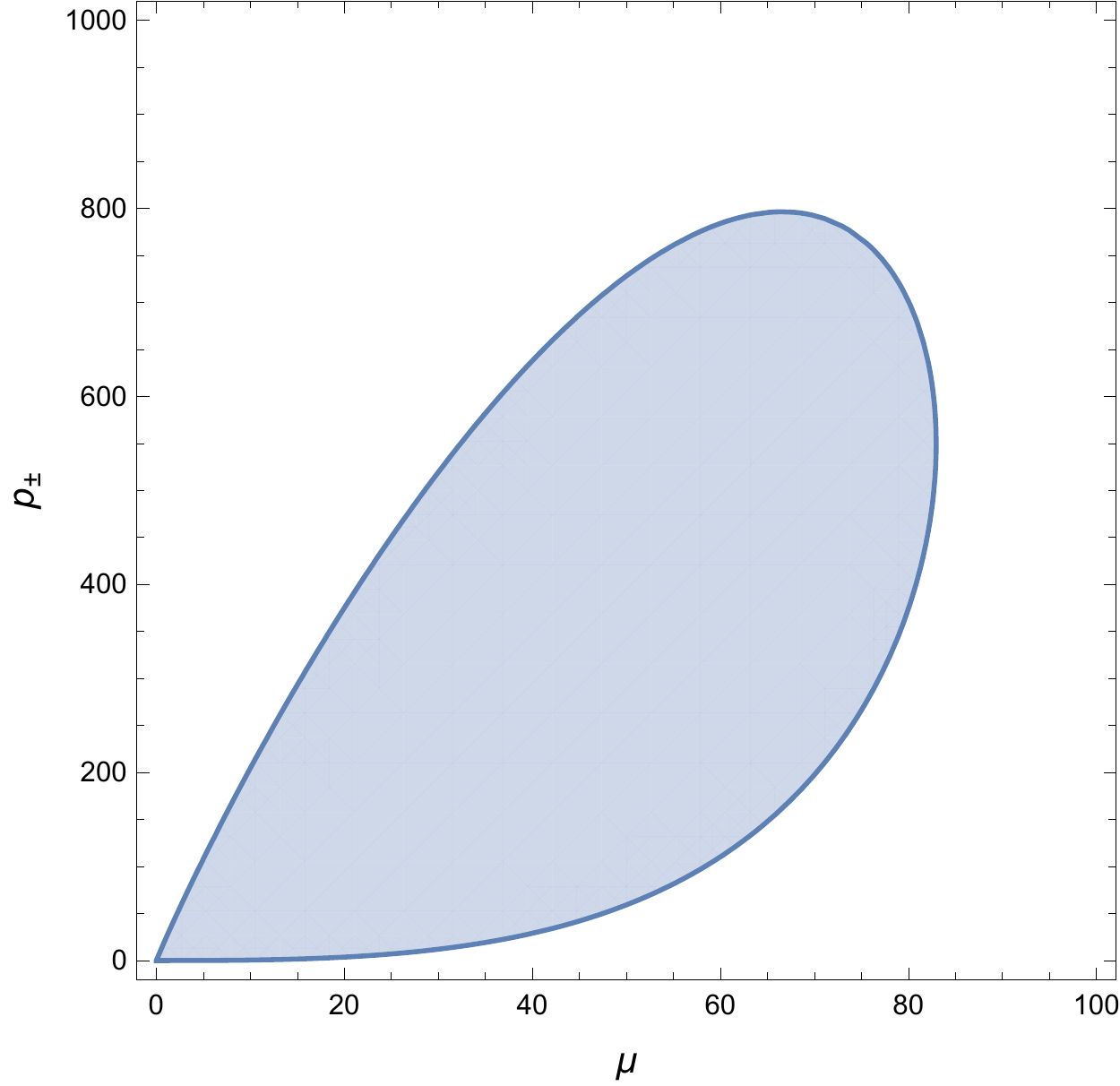}
\includegraphics[scale=.48]{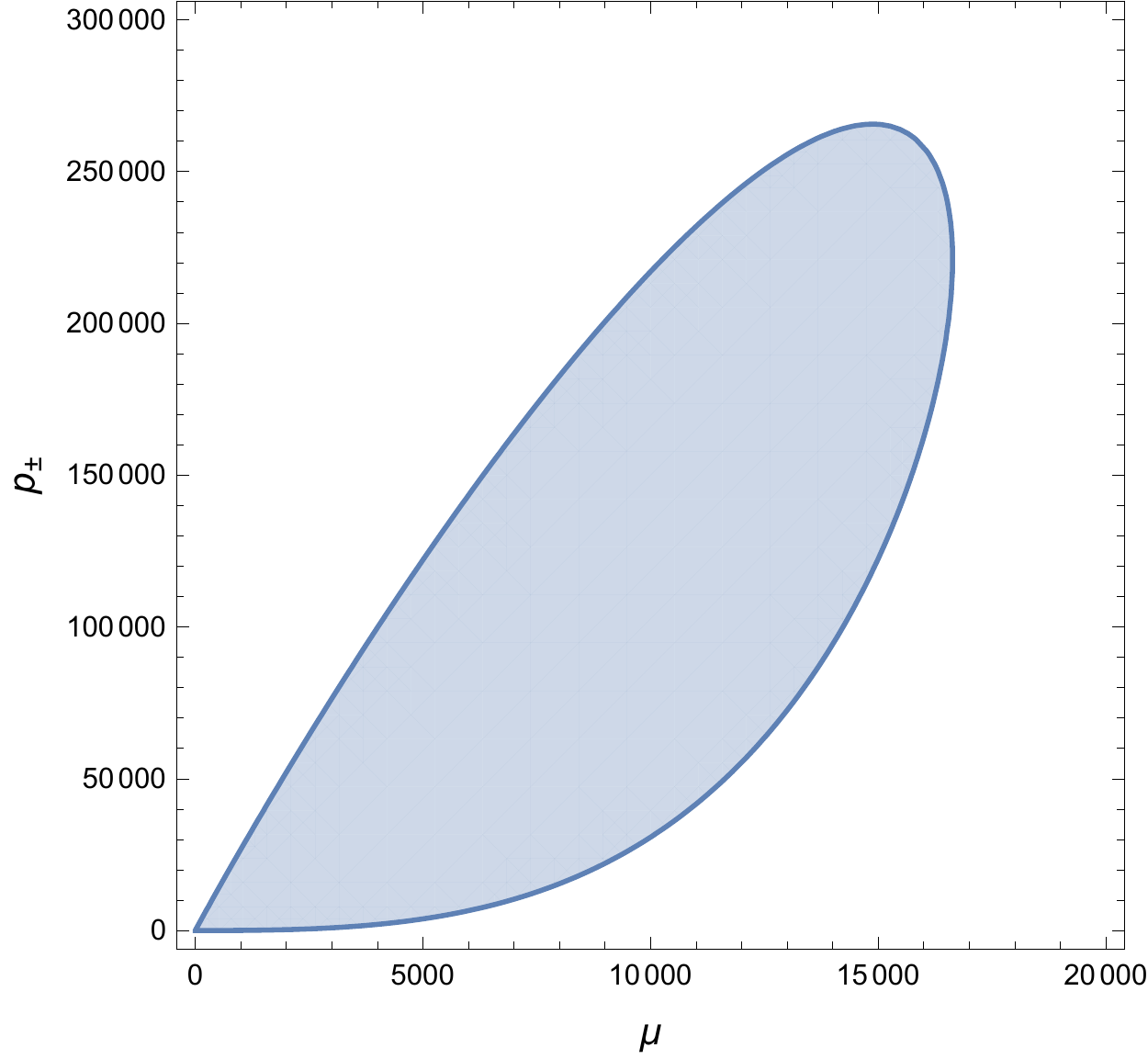}
\caption{\footnotesize The blue region indicates the region of the configuration space $\{\mu,p_{\pm}\}$ in which the condition $\mathcal{V}^{*}_{IX}/\mathcal{K}^{*}< \frac{1}{100}$ is valid, sketched for the three values of the polymer scale $\lambda =0.015,0.0015,0.00015$. The Bianchi IX potential term becomes more and more negligible with the decrease of the polymer scale.}
\label{fig7}
\end{figure*}
\end{widetext}
where the potential term, which accounts for the spatial curvature of the model, reads as\cite{primordial}
\begin{multline}
\label{potanis}
 V_{IX}(\beta _{\pm})= e^{-8\beta _{+}}-4e^{-2\beta _{+}}\cosh (2\sqrt{3}\beta _{-})+ \\ +2e^{4\beta _{+}}\left[ \cosh(4\sqrt{3}\beta_{-})-1 \right].
\end{multline}
Looking at the Eq.(\ref{hamiltoniana BIX}) it is evident that the difference between the Bianchi I model and the Bianchi IX model is the presence of the potential term $ \frac{12 \pi^{2}\hbar^{2}}{l_{p}^{4}}aV_{IX}(\beta_{\pm})$. Being the potential term associated to the anisotropies, it formally enters, performing a WKB expansion in $\hbar$ with a wave function of the Universe \textit{a la} Vilenkin, in the first-order equation, \textit{i.e.} the Schrodinger equation. 
Keeping this in mind, a possible way to estimate the importance of the potential term is to individuate the existence of a particular set of parameters for which the potential term of the Bianchi IX model is negligible with respect to $\frac{p_{+}^{2} + p_{-}^{2}}{a^{3}}$, in other words the kinetic term of the anisotropies. Finding such a regime means that the results for the Bianchi I model obtained in Section \ref{sec:4} can be extended also to the Bianchi IX model and moreover, through the BKL conjecture, to a generic cosmological solution.
To this aim, we consider the ratio between the potential term $\mathcal{V}^{*}_{IX}=\frac{12 \pi^{2}\hbar^{2}}{l_{p}^{4}}a_{min}V_{IX}(\beta_{\pm})$ and the kinetic term related to the anisotropies $\mathcal{K}^{*}=\frac{p_{+}^{2} + p_{-}^{2}}{a_{min}^{3}}$, both evaluated at the bounce.

In FIG. \ref{fig7} are represented, for different values of the polymer scale $\lambda$, the regions in the space of the parameters $\{ \mu, p_{\pm} \}$ where the ratio $\mathcal{V}^{*}_{IX}/\mathcal{K}^{*}$ become not relevant. In particular, the blue regions represent the values of $\{ \mu, p_{\pm} \}$ for which the condition $\mathcal{V}^{*}_{IX}/\mathcal{K}^{*} < \frac{1}{100}$ is valid. 
Therefore, as it is clear from the figure, for any value of the parameter $\lambda$ it is always possible to individuate a non-zero region where the Bianchi IX potential term is negligible with respect to the kinetic term. Furthermore, the blue region becomes bigger as the parameter $\lambda$ becomes smaller. This means that choosing smaller $\lambda$ values implies that the condition for neglecting the potential term with respect to the kinetic term is verified for a large number of parameters couples $\{ \mu, p_{\pm} \}$.
The identification of such a regions bring us to conclude, providing proper parameters in order to neglect the potential term, that the Bianchi IX model in presence of a stiff matter contribution in the polymer approach possesses the same qualitatively features of the Bianchi I model previously analyzed.

This considerations suggest that in the present representation the Bianchi IX model is chaos free as also argued in a LQC approach discussed in \cite{boj1}. Indeed, our approach and the LQC have a significant contact point in the discretization of the Universe volume: here this is due to a polymer paradigm for the discretization of the isotropic coordinate while in LQC the same issue comes from the discrete spectrum of the volume operator even in a more general approach, as prescribed by the general Loop Quantum Gravity approach.

\section{Concluding Remarks}

In this paper, we analyzed in some detail how the anisotropies of a Bianchi type I 
model, represented by the Misner variables $\beta_+$ and $\beta _-$,
behave when a Big-Bounce scenario is inferred via a polymer approach 
to the corresponding Misner variable $\alpha$, describing the 
Universe volume. In order to be able to construct a proper dynamical 
Hilbert space for the anisotropy variables, we adopted a Vilenkin 
WKB representation of the Wheeler-DeWitt equation, in which the 
Universe volume is a quasi-classical configurational coordinate. 
In such a scheme, $\alpha$ recover its most genuine meaning of 
a time-like variable, since it does not correspond to a physical 
degree of freedom of the gravitational cosmological field. 
Since we adopted a polymer quantization procedure for such 
semi-classical component, we de facto deal with a modified classical 
Hamiltonian describing its evolution in the presence of 
stiff matter,\textit{ i.e.} a modified Friedmann equation for the 
presence of a typical length scale of cut-off. 

In the analysis above, we obtained to main relevant achievements: 
i) the anisotropies of the Universe remain, in a Bianchi I model, finite across the 
Big-Bounce and they approach a localized quasi-classical 
behavior, according to the original idea of Misner \cite{misner};

ii) the same behavior remains valid for a Bianchi IX model, since 
for a non-zero set of initial conditions, the potential term, 
due to the spatial curvature (absent in the Bianchi I model) is dynamically
negligible.

The first result suggests that the deviation of a primordial Universe 
from the isotropy can be controlled via the Cauchy problem, 
when the Bounce picture is recovered for the Universe volume. 
The second achievement allows to extend such an intriguing 
primordial feature, from a flat homogeneous Universe to a 
positive curved one. Furthermore, the Bianchi IX model 
has an high degree of generality 
and it mimics the generic cosmological solution near the initial singularity \cite{BKL1982},\cite{primordial}.
By other words, we can infer that the proposed scenario remains 
valid even when we address the dynamics of a generic inhomogeneous 
model, near the Big-Bounce, as ensured by the polymer 
treatment of the $\alpha$ variable. Such a conjecture 
is based on the so-called \emph{long wavelength approximation}, 
according to which, each space point, de facto each causal 
region of the Universe,dynamically decouples near enough to the 
initial singularity\cite{BKL1982},\cite{kirillov1993},\cite{montani1995},\cite{primordial}, here replaced by a Big-Bounce .

From a cosmological point of view, the present study has 
the merit to demonstrate how, in the presence of a cut-off 
physics and a proper interpretation of the Universe wavefunction, 
the anisotropies do not explode asimptotically to the 
primordial turning point and the scenario of a Big-Bounce 
cosmology makes the quasi-isotropic Universe a more general 
solution with respect to a pure classical Einsteinian cosmology.

\end{document}